\documentclass[
 reprint,
 superscriptaddress,
 nofootinbib,
 amsmath,amssymb,
 aps,
 prd,
]{revtex4-2}
\usepackage[colorlinks=true,pdfstartview=FitV,linkcolor=blue,citecolor=blue,urlcolor=blue, hyperfootnotes=true]{hyperref}
\usepackage{wasysym}
\usepackage[separate-uncertainty,retain-explicit-plus,per-mode=symbol,binary-units]{siunitx}
\usepackage{array,mathtools,amssymb,dcolumn}
\usepackage{bm}
\usepackage[below]{placeins}

\usepackage[dvipsnames, table]{xcolor}
\usepackage{tikz}
\usepackage{multirow}
\usepackage{afterpage}
\usepackage{lineno}
\usepackage{paralist}
\usepackage{listings}
\usepackage{array}
\usepackage{cancel}
\usepackage{xspace}

\usepackage{stmaryrd}
\usepackage[version=4]{mhchem}

\usepackage{calc}
\newcommand{\instr}[1]{{\color{Black}#1}} %

\newcommand{\etal}{\emph{et al.}}

\newcommand{\wimpy}{\texttt{WIMpy\_NREFT}}
\newcommand{\chiraleftdm}{\texttt{CHIRALEFT4DM}}

\newcommand{\E}[1]{\ensuremath{\times 10^{#1}}}
\newcommand{\CO}{\mathcal{O}}
\DeclareSIUnit\c{\mbox{$c$}}
\DeclareSIUnit\week{w}
\DeclareSIUnit\year{yr}
\DeclareSIUnit\yr{yr}
\DeclareSIUnit\yr{yr}
\DeclareSIUnit\standard{std}
\DeclareSIUnit\str{sr}
\DeclareSIUnit\ppm{ppm}
\DeclareSIUnit\ppb{ppb}
\DeclareSIUnit\ppt{ppt}
\DeclareSIUnit\pe{PE}
\DeclareSIUnit\spe{SPE}
\DeclareSIUnit\ev{events}
\DeclareSIUnit\hit{hit}
\DeclareSIUnit\hits{hits}
\DeclareSIUnit\bin{(\mbox{5-PE}~bin)}
\DeclareSIUnit\sgm{\mbox{$\sigma$}}
\DeclareSIUnit\rms{RMS}
\DeclareSIUnit\keVr{\mbox{keV$_{\rm nr}$}}
\DeclareSIUnit\keVee{\mbox{keV$_{\rm ee}$}}
\DeclareSIUnit\ph{photons}
\DeclareSIUnit\pm{PMT}
\DeclareSIUnit\inch{''}
\DeclareSIUnit\bit{bit}

\DeclareSIUnit\sample{S}
\DeclareSIUnit\barn{b}
\DeclareSIUnit\bara{bar}
\DeclareSIUnit\Curie{Ci}
\DeclareSIUnit\psi{psi}
\DeclareSIUnit{\msun}{\mbox{M$_\odot$}}
\DeclareSIUnit\mK{\milli\kelvin}
\DeclareSIUnit\micron{\micro\metre}
\DeclareSIUnit\liveday{\mbox{live-days}}
\DeclareSIUnit\tonneday{\mbox{tonne$\cdot$day}}
\DeclareSIUnit\days{\mbox{days}}

\usepackage[symbol*]{footmisc}
\DefineFNsymbolsTM{otherfnsymbols}{%
  \textdagger    \dagger
  \textasteriskcentered *
  \textbardbl    \|%
  \textparagraph \mathparagraph
  \textbullet \circ
  \textdaggerdbl \ddagger
}%
\setfnsymbol{otherfnsymbols}

\newcommand{\PaperTwoWIMPLimitOneHundredGeV}{\SI{3.9E-45}{\square\cm}}
\newcommand{\PaperTwoWIMPLimitOneTeV}{\SI{1.5E-44}{\square\cm}}

\newcommand{\SnolabDepth}{\SI{2}{\km}}

\newcommand{\larmasserror}{\SI{3279 \pm 96}{\kg}}

\newcommand{\LGLength}{\SI{45}{\cm}}

\newcommand{\AVThickness}{\SI{5}{\cm}}

\newcommand{\PaperTwoFiducialMass}{\instr{\SI{824\pm25}{\kg}}}

\newcommand{\eftlink}{\url{https://zenodo.org/record/3998892}}

\newcommand{\PaperTwoLiveTimeNum}{231}

\newcommand{\PaperTwoLiveTime}{\SI{\PaperTwoLiveTimeNum}{\liveday}}
\newcommand{\PaperTwoExpoNum}{758}
\newcommand{\PaperTwoExpo}{\SI{\PaperTwoExpoNum}{\tonneday}}
 
\newcommand{\arnine}{\mbox{$^{39}$Ar}}
\newcommand{\arforty}{\mbox{$^{40}$Ar}}

\newcommand{\AArArThreeNineActivity}{\SI{0.95\pm0.05}{\becquerel\per\kg}}

\newcommand{\Gaia}{\mbox{\textit{Gaia}}\xspace}
\newcommand{\SSDSGaia}{\mbox{SDSS-\textit{Gaia}}}
\newcommand{\GaiaSausage}{\mbox{\Gaia\ Sausage}\xspace}
\newcommand{\SNOLAB}{\mbox{SNOLAB}}
\newcommand{\DEAP}{\mbox{DEAP-3600}}
\newcommand{\SHM}{\mbox{SHM}}
\newcommand{\AV}{\mbox{AV}}

\newcommand{\LGs}{\mbox{LGs}}

\newcommand{\CRESST}{\mbox{CRESST}}
\newcommand{\SCDMS}{\mbox{SuperCDMS}}
\newcommand{\XenonH}{\mbox{XENON100}}
\newcommand{\Xenon}{\mbox{XENON1T}}

\newcommand{\DSf}{\mbox{DarkSide-50}}

\newcommand{\PMT}{\mbox{PMT}}

\newcommand{\PMTs}{\mbox{PMTs}}

\newcommand{\WIMP}{\mbox{WIMP}}

\newcommand{\WIMPs}{\mbox{WIMPs}}

\newcommand{\LXe}{\ce{LXe}}
\newcommand{\LAr}{\ce{LAr}}

\newcommand{\vperpsquare}{\mbox{$v^2_\perp$}}

\newcommand{\CL}{C.~L.}

\newcommand{\SubstructureFraction}{\mbox{$\eta_\text{sub}$}}
\newcommand{\WIMPMassSymbol}{\mbox{$m_\chi$}}

\newcommand{\WIMPMassFortyGeV}{\SI{40}{\GeV\per\square\c}}

\newcommand{\WIMPMassHundredGeV}{\SI{100}{\GeV\per\square\c}}

\newcommand{\WIMPMassThreeTeV}{\SI{3}{\TeV\per\square\c}}

\newcommand{\NinetyPerCentCL}{\mbox{\SI{90}{\percent}~\CL}}

\newcommand{\NR}{\mbox{NR}}
\newcommand{\NRs}{\mbox{NRs}}
\newcommand{\ER}{\mbox{ER}}
\newcommand{\ERs}{\mbox{ERs}}

\newcommand{\alpps}{\mbox{$\alpha$ particles}}

\newcommand{\alpds}{\mbox{$\alpha$-decays}}

\newcommand{\betds}{\mbox{$\beta$-decays}}

\newcommand{\PSD}{\mbox{PSD}}
\newcommand{\FPrompt}{\mbox{F$_{\text{prompt}}$}}

\usepackage{graphicx}
\usepackage{dcolumn}

\usepackage{bm}
\usepackage{makecell}
\usepackage{cellspace} 
\usepackage{subfigure} 
\usepackage{float}
\usepackage[shortlabels]{enumitem}

\newcommand{\vperp}{\ensuremath{{v_\perp}}}
\newcommand{\vvperp}{\ensuremath{\vec{v}_\perp}}

\begin{document}
\preprint{APS/123-QED}
\title{Constraints on dark matter-nucleon effective couplings in the presence of kinematically distinct halo substructures using the DEAP-3600 detector}

\newcommand{\UofA}{Department of Physics, University of Alberta, Edmonton, Alberta, T6G 2R3, Canada}
\newcommand{\AsC}{AstroCeNT, Nicolaus Copernicus Astronomical Center, Polish Academy of Sciences, Rektorska 4, 00-614 Warsaw, Poland}
\newcommand{\CNL}{Canadian Nuclear Laboratories, Chalk River, Ontario, K0J 1J0, Canada}
\newcommand{\CIEMAT}{Centro de Investigaciones Energ\'eticas, Medioambientales y Tecnol\'ogicas, Madrid 28040, Spain}
\newcommand{\CU}{Department of Physics, Carleton University, Ottawa, Ontario, K1S 5B6, Canada}
\newcommand{\LNGSA}{INFN Laboratori Nazionali del Gran Sasso, Assergi (AQ) 67100, Italy}
\newcommand{\RHUL}{Royal Holloway University London, Egham Hill, Egham, Surrey TW20 0EX, United Kingdom}
\newcommand{\LU}{Department of Physics and Astronomy, Laurentian University, Sudbury, Ontario, P3E 2C6, Canada}
\newcommand{\UNAM}{Instituto de F\'isica, Universidad Nacional Aut\'onoma de M\'exico, A.\,P.~20-364, M\'exico D.\,F.~01000, M\'exico}
\newcommand{\INFN}{INFN Napoli, Napoli 80126, Italy}
\newcommand{\PRISMA}{PRISMA$^+$, Cluster of Excellence and Institut f\"ur Kernphysik, Johannes Gutenberg-Universit\"at Mainz, 55128 Mainz, Germany}
\newcommand{\PU}{Physics Department, Princeton University, Princeton, NJ 08544, USA}
\newcommand{\QU}{Department of Physics, Engineering Physics, and Astronomy, Queen's University, Kingston, Ontario, K7L 3N6, Canada}
\newcommand{\RAL}{Rutherford Appleton Laboratory, Harwell Oxford, Didcot OX11 0QX, United Kingdom}
\newcommand{\SL}{SNOLAB, Lively, Ontario, P3Y 1N2, Canada}
\newcommand{\Sussex}{University of Sussex, Sussex House, Brighton, East Sussex BN1 9RH, United Kingdom}
\newcommand{\TRIUMF}{TRIUMF, Vancouver, British Columbia, V6T 2A3, Canada}
\newcommand{\TUM}{Department of Physics, Technische Universit\"at M\"unchen, 80333 Munich, Germany}
\newcommand{\Napoli}{Physics Department, Universit\`a degli Studi ``Federico II'' di Napoli, Napoli 80126, Italy}
\newcommand{\LBLNSD}{Currently: Nuclear Science Division, Lawrence Berkeley National Laboratory, Berkeley, CA 94720}
\newcommand{\kurchatov}{National Research Centre Kurchatov Institute, Moscow 123182, Russia}
\newcommand{\MEPhI}{National Research Nuclear University MEPhI, Moscow 115409, Russia}
\newcommand{\MI}{Arthur B. McDonald Canadian  Astroparticle Physics Research Institute, Queen's University, Kingston ON K7L 3N6,Canada}
\newcommand{\PI}{Perimeter Institute for Theoretical Physics, Waterloo ON N2L 2Y5, Canada}
\newcommand{\UdSCag}{Physics Department, Universit\`a degli Studi di Cagliari, Cagliari  09042, Italy}
\newcommand{\INFNCag}{INFN Cagliari, Cagliari 09042, Italy}
\newcommand{\NCNR}{BP2, National Centre for Nuclear Research, ul. Pasteura 7, 02-093 Warsaw, Poland}

\affiliation{\UofA}
\affiliation{\AsC}
\affiliation{\CNL}
\affiliation{\CIEMAT}
\affiliation{\CU}
\affiliation{\Napoli}
\affiliation{\UdSCag}
\affiliation{\LNGSA}
\affiliation{\LU}
\affiliation{\UNAM}
\affiliation{\kurchatov}
\affiliation{\MEPhI}
\affiliation{\INFN}
\affiliation{\INFNCag}
\affiliation{\NCNR}
\affiliation{\PRISMA}
\affiliation{\PU}
\affiliation{\QU}
\affiliation{\RHUL}
\affiliation{\SL}
\affiliation{\Sussex}
\affiliation{\TRIUMF}
\affiliation{\TUM}
\affiliation{\MI}
\affiliation{\PI}

\author{P.~Adhikari}\affiliation{\CU} 
\author{R.~Ajaj}\affiliation{\CU}\affiliation{\MI}
\author{D.\,J.~Auty}\affiliation{\UofA}
\author{C.\,E.~Bina}\affiliation{\UofA}\affiliation{\MI}
\author{W.~Bonivento}\affiliation{\INFNCag}
\author{M.\,G.~Boulay}\affiliation{\CU}
\author{M.~Cadeddu}\affiliation{\UdSCag}\affiliation{\INFNCag}
\author{B.~Cai}\affiliation{\CU}\affiliation{\MI} 
\author{M.~C\'ardenas-Montes}\affiliation{\CIEMAT}
\author{S.~Cavuoti}\affiliation{\Napoli}\affiliation{\INFN}
\author{Y.~Chen}\affiliation{\UofA}
\author{B.\,T.~Cleveland}\affiliation{\SL}\affiliation{\LU}
\author{J.\,M.~Corning}\affiliation{\QU} 
\author{S.~Daugherty}\affiliation{\LU} 
\author{P.~DelGobbo}\affiliation{\CU}\affiliation{\MI} 
\author{P.~Di~Stefano}\affiliation{\QU} 
\author{L.~Doria}\affiliation{\PRISMA} 
\author{M.~Dunford}\affiliation{\CU}
\author{A.~Erlandson}\affiliation{\CU}\affiliation{\CNL}
\author{S.\,S.~Farahani}\affiliation{\UofA} 
\author{N.~Fatemighomi}\affiliation{\SL}\affiliation{\RHUL}
\author{G.~Fiorillo}\affiliation{\Napoli}\affiliation{\INFN}
\author{D.~Gallacher}\affiliation{\CU}
\author{E.\,A.~Garc{\'e}s}\affiliation{\UNAM} 
\author{P.~Garc\'ia~Abia}\affiliation{\CIEMAT}
\author{S.~Garg}\affiliation{\CU}
\author{P.~Giampa}\affiliation{\TRIUMF}
\author{D.~Goeldi}\affiliation{\CU}\affiliation{\MI}
\author{P.~Gorel}\affiliation{\SL}\affiliation{\LU}\affiliation{\MI}
\author{K.~Graham}\affiliation{\CU}
\author{A.~Grobov}\affiliation{\kurchatov}\affiliation{\MEPhI} 
\author{A.\,L.~Hallin}\affiliation{\UofA}
\author{M.~Hamstra}\affiliation{\CU}
\author{T. Hugues}\affiliation{\AsC} 
\author{A.~Ilyasov}\affiliation{\kurchatov}\affiliation{\MEPhI} 
\author{A.~Joy}\affiliation{\UofA}\affiliation{\MI}
\author{B.~Jigmeddorj}\affiliation{\CNL}
\author{C.\,J.~Jillings}\affiliation{\SL}\affiliation{\LU}
\author{O.~Kamaev}\affiliation{\CNL}
\author{G.~Kaur}\affiliation{\CU}
\author{A.~Kemp}\affiliation{\RHUL}
\author{I.~Kochanek}\affiliation{\LNGSA}
\author{M.~Ku{\'z}niak}\affiliation{\AsC}\affiliation{\CU}\affiliation{\MI}
\author{M. Lai}\affiliation{\UdSCag}\affiliation{\INFNCag}
\author{S.~Langrock}\affiliation{\LU}\affiliation{\MI}
\author{B.~Lehnert}\altaffiliation{\LBLNSD}\affiliation{\CU}
\author{N.~Levashko}\affiliation{\kurchatov}\affiliation{\MEPhI} 
\author{X.~Li}\affiliation{\PU}
\author{O.~Litvinov}\affiliation{\TRIUMF}
\author{J.~Lock}\affiliation{\CU}
\author{G.~Longo}\affiliation{\Napoli}\affiliation{\INFN}
\author{I.~Machulin}\affiliation{\kurchatov}\affiliation{\MEPhI} 
\author{A.\,B.~McDonald}\affiliation{\QU}
\author{T.~McElroy}\affiliation{\UofA}
\author{J.\,B.~McLaughlin}\affiliation{\RHUL}
\author{C.~Mielnichuk}\affiliation{\UofA}
\author{J.~Monroe}\affiliation{\RHUL}
\author{G.~Olivi\'ero}\affiliation{\CU}\affiliation{\MI} 
\author{S.~Pal}\affiliation{\UofA}\affiliation{\MI} 
\author{S.\,J.\,M.~Peeters}\affiliation{\Sussex}
\author{V.~Pesudo}\affiliation{\CIEMAT}
\author{M.-C.~Piro}\affiliation{\UofA}\affiliation{\MI}
\author{T.\,R.~Pollmann}\affiliation{\TUM}
\author{E.\,T.~Rand}\affiliation{\CNL}
\author{C.~Rethmeier}\affiliation{\CU}
\author{F.~Reti\`ere}\affiliation{\TRIUMF}
\author{I. Rodr\'iguez-Garc\'ia}\affiliation{\CIEMAT}
\author{L.~Roszkowski}\affiliation{\AsC}\affiliation{\NCNR}
\author{E.~Sanchez~Garc\'ia}\affiliation{\CIEMAT}
\author{T.~S\'anchez-Pastor}\affiliation{\CIEMAT} 
\author{R.~Santorelli}\affiliation{\CIEMAT}
\author{D.~Sinclair}\affiliation{\CU}
\author{P.~Skensved}\affiliation{\QU}
\author{B.~Smith}\affiliation{\TRIUMF}
\author{N.\,J.\,T.~Smith}\affiliation{\SL}\affiliation{\LU}
\author{T.~Sonley}\affiliation{\SL}\affiliation{\MI}
\author{R.~Stainforth}\affiliation{\CU}
\author{M.~Stringer}\affiliation{\QU}\affiliation{\MI} 
\author{B.~Sur}\affiliation{\CNL}
\author{E.~V\'azquez-J\'auregui}\affiliation{\UNAM}\affiliation{\LU}
\author{S.~Viel}\affiliation{\CU}\affiliation{\MI}
\author{A.\,C.~Vincent}\affiliation{\QU}\affiliation{\MI}\affiliation{\PI}
\author{J.~Walding}\affiliation{\RHUL}
\author{M.~Waqar}\affiliation{\CU}\affiliation{\MI}
\author{M.~Ward}\affiliation{\QU}
\author{S.~Westerdale}\affiliation{\INFNCag}\affiliation{\CU}
\author{J.~Willis}\affiliation{\UofA}
\author{A.~Zu\~niga-Reyes}\affiliation{\UNAM}
\collaboration{DEAP Collaboration}\email{deap-papers@snolab.ca}\noaffiliation

\date{\today}

\begin{abstract}
\DEAP\ is a single-phase liquid argon detector aiming to directly detect Weakly Interacting Massive Particles (\WIMPs), located at \SNOLAB\ (Sudbury, Canada).
After analyzing data taken during the first year of operation, a null result was used to place an upper bound on the \WIMP-nucleon spin-independent, isoscalar cross section.
This study reinterprets this result within a Non-Relativistic Effective Field Theory framework, and further examines how various possible substructures in the local dark matter halo may affect these constraints. Such substructures are hinted at by kinematic structures in the local stellar distribution observed by the \Gaia\ satellite and other recent astronomical surveys. These include the \GaiaSausage\ (or Enceladus), as well as a number of distinct streams identified in recent studies.
Limits are presented for the coupling strength of the effective contact interaction operators $\CO_1$, $\CO_3$, $\CO_5$, $\CO_8$, and $\CO_{11}$, considering isoscalar, isovector, and xenonphobic scenarios, as well as the specific operators corresponding to millicharge, magnetic dipole, electric dipole, and anapole interactions.
The effects of halo substructures on each of these operators are explored as well, showing that the $\CO_5$ and $\CO_8$  operators are particularly sensitive to the velocity distribution, even at dark matter masses above \SI{100}{\GeV\per\square\c}. 

\end{abstract}
\maketitle

\section{\label{sec:intro}Introduction}

An abundance of astrophysical and cosmological observations indicate that the majority of the matter in the universe is comprised of non-baryonic ``dark matter'' (DM)~\cite{planck_collaboration_planck_2018,Bertone:2005bi}.
Despite this evidence, there have been no unambiguous direct or indirect detection signals of DM interacting with the Standard Model, and the particle nature of DM is still unknown.
One promising candidate is the Weakly Interacting Massive Particle (\WIMP)~\cite{Feng:2010dz}, which may couple to nucleons at the weak scale or below and have a mass on the order of \SI{100}{\GeV\per\square\c}.
Such particles are predicted to produce low-energy ($\lesssim$\SI{100}{\keV}) nuclear recoils (\NRs) on target nuclei, allowing direct detection experiments to constrain the \WIMP-nucleon coupling strength~\cite{goodman_detectability_1985}.

The predicted rate $R$ of observed DM particles scattering in a detector to produce a recoiling target nucleus of energy $E_R$ is given by
\begin{equation}
    \frac{dR}{dE_R} = \frac{\rho_T}{m_T}\frac{\rho_\chi}{m_\chi} \varepsilon(E_R)\int^\infty_{v_\text{min}}vf_\chi^{\varoplus}(\vec{v})\frac{d\sigma}{dE_R}d^3\vec{v}
    \label{eq:diffscat}
\end{equation}
where $\rho_T$ is the density of the target nucleus with nuclear mass $m_T$, $\rho_\chi$ is the density of the DM with mass \WIMPMassSymbol, $f^\varoplus_\chi(\vec{v})$ is the Earth-frame velocity distribution of the DM, $d\sigma/dE_R$ is the differential scattering cross section, $v_\text{min}$ is the minimum DM velocity that can produce a recoil of energy $E_R$, and $\varepsilon(E_R)$ is the efficiency for detecting NRs of energy $E_R$.
This equation can be used to predict the number of events a direct detection experiment would expect to see from a given DM model, which can then be used to constrain such models.

This paper builds upon the analysis of \DEAP\ data presented in~\cite{Ajaj:2019imk}, in which a \PaperTwoExpo\ total exposure was collected with \PaperTwoLiveTime\ over the course of one year. No \WIMP-like events were observed in this data set.
From these results, \DEAP\ placed leading constraints on elastic \WIMP-nucleon scattering with an argon target, excluding cross sections above \PaperTwoWIMPLimitOneHundredGeV\ (\PaperTwoWIMPLimitOneTeV) for \WIMP\ masses of \SI{100}{\GeV\per\square\c} (\SI{1}{\TeV\per\square\c}).
These limits assume the Standard Halo Model (\SHM) specified in~\cite{mccabe_astrophysical_2010} and hold for isoscalar, spin-independent \WIMP-nucleon interactions, with a massive mediator described by a simple constant contact cross section. 
The present analysis investigates how variations on these assumptions, which particularly affect $f^\varoplus_\chi(\vec{v})$ and $d\sigma/dE_R$ in Eq.~\eqref{eq:diffscat}, impact constraints on DM-nucleon interactions.

Recent observational and theoretical developments have suggested that these standard descriptions of the DM halo and particle interactions may be oversimplified, and can miss important features, or risk misidentifying or overconstraining a potential signal. 

The European Space Agency's \Gaia\ space mission was launched in 2013 with the goal of measuring the positions and velocities of a billion Milky Way stars with unprecedented astrometric precision. 
Between \Gaia's second data release \cite{gaia_collaboration_gaia_2018} and data from the Sloan Digital Sky Survey (SDSS)~\cite{york_sloan_2000}, a number of groups have identified rich kinematic substructure in the local stellar distribution, beyond the expected halo and disk stars. 
A local, anisotropic component dubbed the ``\GaiaSausage'' \cite{10.1093/mnras/sty982,Myeong_2018,myeong_sausage_2018}, or ``\Gaia\ Enceladus'' \cite{2018Natur.563...85H}, has been robustly identified by several groups as a likely remnant of a merger event with a massive dwarf galaxy. Smaller, low- and high-velocity clumps, shards, and streams have also been characterized in the \Gaia\ and SDSS data, as well as in prior surveys~\cite{Myeong:2017skt,10.1093/mnras/sty1403,OHare:2018trr,Koppelman_2018,OHare:2019qxc,Necib:2019zbk,Necib:2019zka,koppelman_characterization_2019}. 
While the direct implications of these smaller structures on the local \textit{dark matter} distribution is still debated \cite{Bozorgnia:2018pfa,Bozorgnia:2019mjk}, it is widely accepted that an association with the DM could imply important modifications of the expected signal at direct detection experiments~\cite{Helmi:2008eq,baushev_extragalactic_2013,freese_direct_2001,stiff_signatures_2001,gelmini_weakly_2001,green_potential_2001,sikivie_secondary_1997,Lisanti:2011as,evans_shm:_2018,OHare:2018trr,necib_under_2019,OHare:2019qxc,Ibarra:2019jac}. 
Considering these uncertainties, the present work explores constraints based on possible DM phase space substructures correlated with observed stellar structures in the solar neighborhood.

Theoretical developments throughout the past decade have also highlighted the importance of considering DM-nucleon interactions beyond the standard spin-independent and spin-dependent interactions. 
Two-to-two interactions generically yield cross sections that depend on the Lorentz-invariant Mandelstam $s$, $t$ and $u$ kinematic quantities. 
In the non-relativistic elastic-scattering limit, these interactions translate into a dependence on the relative velocity and transferred momentum. 
While many SUSY models predict scattering amplitudes that are dominated by a constant term, cancellations or long-range forces can allow terms that vary with momentum transfer or DM velocity to dominate.
Such interactions may also dominate under more general frameworks, depending on the nature of the mediator.
These interactions can be generically parametrized in terms of a now-standard set of non-relativistic effective operators (NREOs, also called NREFT operators) \cite{Fan:2010gt,fitzpatrick_effective_2013,Catena:2015uha,Hoferichter:2018acd}, for which the nuclear scattering cross sections depend on exchanged momentum, relative velocity, as well as nucleon and DM spins, and isospin coupling. 
Because nuclei couple differently to different operators, such an approach can highlight the complementarity between different detector techniques and materials \cite{Gluscevic:2015sqa}. 
For instance, certain (isospin-violating) combinations of proton and neutron couplings can lead to a suppressed signal in xenon \cite{Feng:2013vod,Gresham:2013mua,Yaguna:2019llp}.
Effective operator interactions have previously been considered in analyses of \SCDMS~\cite{Schneck:2015eqa}, \XenonH~\cite{Aprile:2017aas}, \CRESST~\cite{Angloher:2018fcs}, \Xenon~\cite{Aprile:2018cxk}, and \DSf~\cite{darkside-50_collaboration_effective_2020}. 

Recently, the effects of the \GaiaSausage\ on a future xenon-based experiment were examined for several NREOs~\cite{Buch:2019aiw}.
This study showed that the velocity distribution of the \GaiaSausage\ led to lower momentum transfers---and therefore reduced sensitivities, except at higher DM masses, where an increase in recoils below \SIrange{5}{10}{\keV} could yield a slight improvement in sensitivity.

The present work employs \DEAP\ data and considers a broad range of possible isospin properties and mediators, along with the simultaneous effects of potential kinematically distinct halo substructures that may vary from the \SHM.

This article is structured as follows.
Section~\ref{sec:detector} provides a brief description of the detector and the event reconstruction. 
Section~\ref{sec:models} describes the halo substructures and DM-nucleon operators under consideration.
Section~\ref{sec:analysis} details the implementation of the models and analysis.
Section~\ref{sec:results} provides the resulting limits from this analysis, and Section~\ref{sec:conclusion} discusses the implications.

\section{\label{sec:detector}The Detector}

\DEAP\ is a DM direct detection experiment located \SnolabDepth\ underground at \SNOLAB, in Sudbury, Canada. 
The detector is described in detail in~\cite{amaudruz_design_2017}.

The active volume of the detector consists of \larmasserror\ of liquid argon (\LAr), contained in a \AVThickness-thick acrylic vessel (\AV).
This volume is viewed by an array of 255 inward-facing Hamamatsu R5912 HQE low radioactivity photomultiplier tubes (\PMTs), which are separated from the \AV\ by \LGLength\ acrylic light guides (\LGs).
The top of the \AV\ opens to the neck, through which the detector was filled.
The detector sits inside a water tank, which acts as a shield against external radiation and a Cherenkov muon veto.

Pulse shape discrimination (\PSD) forms a cornerstone of the \DEAP\ analysis by separating slower scintillation pulses due to electronic recoils (\ERs) from faster signals induced by \NRs~\cite{amaudruz_measurement_2016,deapPSD2020}. 
\NRs\ may be caused by rarer interactions from neutrons and \alpps\ or by DM scattering on an \arforty\ nucleus. They therefore constitute candidate signal events.
\ERs, on the other hand, constitute the majority of the backgrounds, mostly coming from \betds\ of \arnine, which is naturally present in \DEAP's atmospherically-derived \LAr\ at a concentration of \AArArThreeNineActivity~\cite{calvo_backgrounds_2018,dunford_2018}.
The \ER\ backgrounds are discussed in more detail in~\cite{deap_collaboration_electromagnetic_2019}.

The \PMT\ calibration and characterization, discussed in~\cite{the_deap_collaboration_-situ_2017}, provides input to a photoelectron-counting algorithm, which removes afterpulses, following the method in~\cite{butcher_method_2017,butcher_2015,burghardt_2018}.
This algorithm improves the energy resolution and efficiency of PSD.

\begin{figure}
    \centering
    \includegraphics[width=\linewidth]{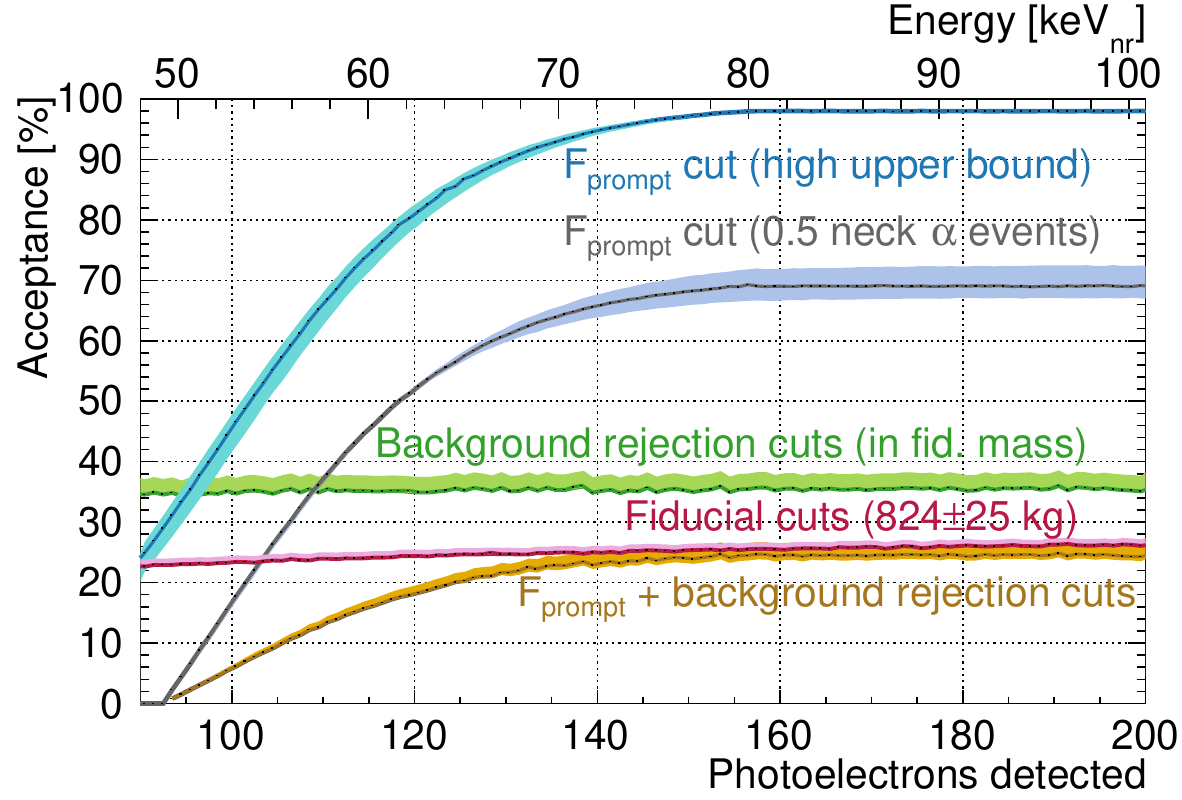}
    \caption{WIMP acceptance, broken down by cut type. The total acceptance is from the \FPrompt +background rejection cuts and the fiducial cuts. From~\cite{Ajaj:2019imk}. } 
    \label{fig:NRaccep}
\end{figure}

The energy region of interest used in this analysis spans the range \SIrange{50}{100}{\keVr}, where \SI{}{\keVr} denotes energy deposited in nuclear recoils.
The NR acceptance in this region is illustrated in Figure~\ref{fig:NRaccep}. ER backgrounds are removed by the \FPrompt\ cut. Backgrounds induced by Cherenkov and \alpds\ in the detector neck are moved by an additional \FPrompt\ cut and the background rejection cuts, and neutron-induced and surface backgrounds are removed by fiducial cuts. After applying PSD and background rejection cuts, the NR acceptance starts at \SI{0}{\percent} at \SI{50}{\keVr} and reaches an approximately constant value near \SI{25}{\percent} above \SI{68}{\keVr}, within the \PaperTwoFiducialMass\ fiducial mass.
Additional details about the analysis are discussed in~\cite{Ajaj:2019imk}.

\section{\label{sec:models}Models}

Variations in astrophysical and particle physics models describing DM are considered in this analysis, as manifest in the $f^\varoplus_\chi(\vec{v})$ and $d\sigma/dE_R$ terms in Eq.~\eqref{eq:diffscat}, respectively. 
Since Eq.~\eqref{eq:diffscat} depends on the product of both terms, simultaneous variations of both models may introduce distinctive behavior.
This section describes the models considered in the present analysis.

\begin{figure*}
    \centering
    \includegraphics[width=\linewidth]{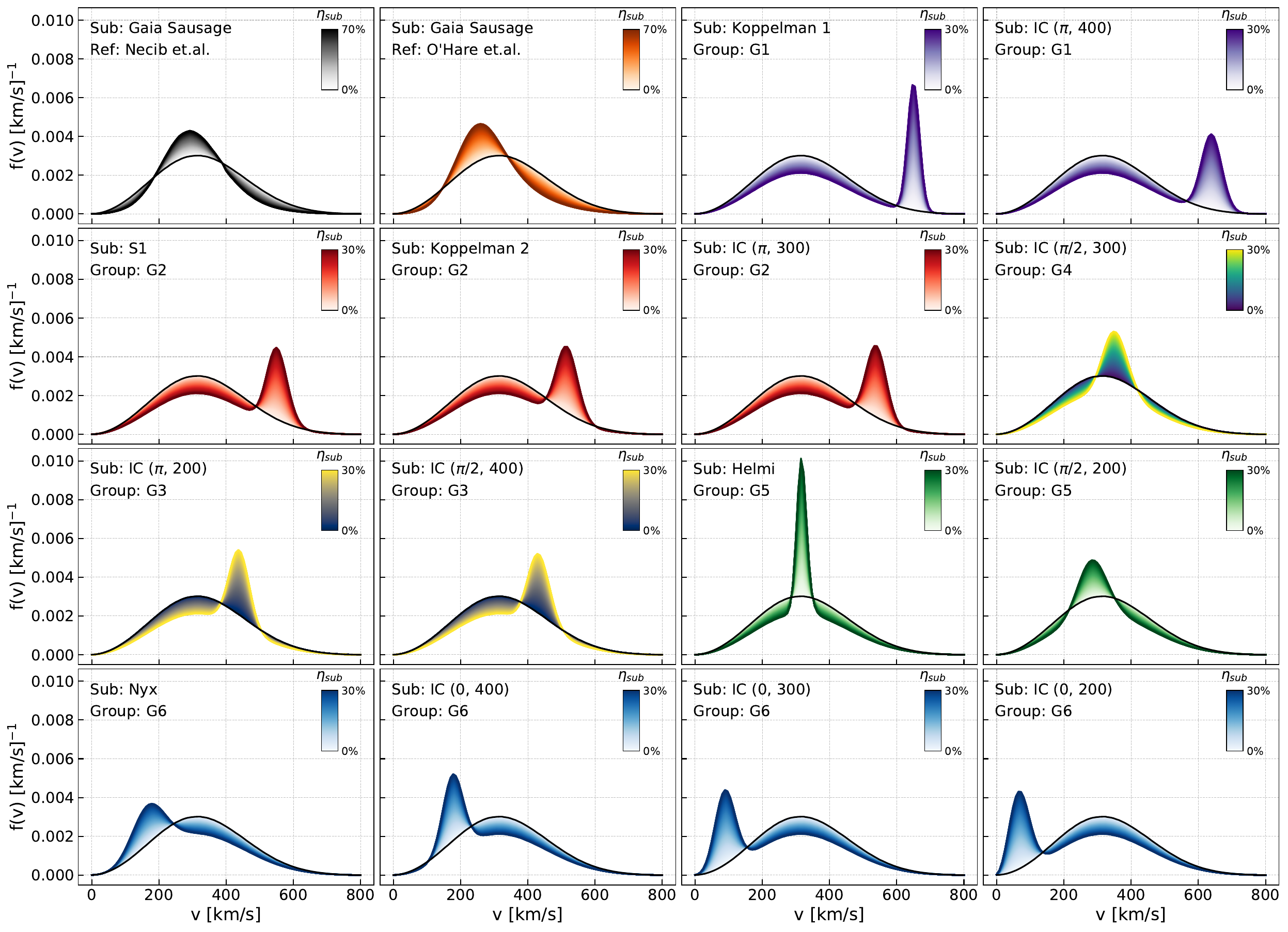}
    \caption{Velocity distributions modeled for this analysis, arranged into groups, including two \GaiaSausage\ models. The first substructure listed in each group marks the chosen representative in Tab.~\ref{tab:vdfs}. The color gradient indicates the relative DM density in each substructure, varying from \SI{0}{\percent} (light) to \SI{30}{\percent} (dark), with the exception of the two \GaiaSausage\ models, which go up to \SI{70}{\percent}. The solid black line corresponds to the SHM.} 
    \label{fig:allvdf}
\end{figure*}

\subsection{\label{subsec:halo}Non-thermal halo components}

The SHM assumes an isotropic thermal distribution for DM in the ``round halo'' of the Milky Way. 
This distribution is described by a Maxwell-Boltzmann distribution with a cutoff at the galactic escape speed, given in the galactic rest frame by,
\begin{equation}
    f^\text{gal}_\text{SHM}(\vec{v}) = N_\text{SHM} \times e^{-\frac{1}{2}|\vec{v}|^2/\sigma_0^2}\times\Theta(v_\text{esc}-|\vec{v}|)
\end{equation}
where $N_\text{SHM}$ is a normalization constant, $\Theta(x)$ is the Heaviside step function, $v_\text{esc}$ is the galactic escape speed, and $\sigma_0$ is the DM velocity dispersion. In the case of an isotropic Maxwellian distribution in a central potential, this is related to the local standard of rest velocity $\vec{v}_0$, via $\sigma_0=|\vec{v}_0|/\sqrt{2}$.
The velocity of the Earth in the galactic rest frame is given by $\vec{v}_\varoplus = \vec{v_0} + \vec{v}_\varoast + \vec{v}_\varoplus^\text{\,sun}$, where $\vec{v}_\varoast$ is the Sun's peculiar velocity and $\vec{v}_\varoplus^\text{\,sun}$ is the velocity of the Earth relative to the Sun. As in Ref.~\cite{OHare:2019qxc}, the value on March 9$^\text{th}$ was chosen, which approximates the time-averaged speed distribution.
The values for these parameters used in this analysis are summarized in Tab.~\ref{tab:shmparams}.
It is worth noting that other authors have suggested modified versions of this model~\cite{mccabe_astrophysical_2010,evans_shm:_2018}.

\begin{table}[htb]
    \centering
    \setcellgapes{2pt}\makegapedcells \renewcommand\theadfont{\normalsize\bfseries}
    \caption{Parameters describing the SHM used in this analysis, denoting the local DM density $\rho_\chi$, Earth's velocity relative to the Sun $\vec{v}_\varoplus^\text{\,sun}$ (chosen as the value on March 9$^\text{th} $ to approximate the time-averaged speed distribution), the modal velocity of the local standard of rest at the Sun's position in the Milky Way $\vec v_0$, the Sun's peculiar velocity $\vec v_\varoast$ with respect to $\vec v_0$, and the escape speed of the Milky Way $ v_{\mathrm{esc}}$, respectively. Vectors are given as $(v_r,v_\theta,v_\phi)$ with $r$ pointing radially inward and $\phi$ in the direction of the Sun's motion.}
    \begin{tabular}{llc}\hline
        Parameter & Value & Ref. \\\hline\hline
         $\rho_\chi$   & \SI{0.3}{\GeV\per\square\c\per\cubic\cm} & \cite{lewin_review_1996} \\ 
         $\vec{v}_\varoplus^\text{\,sun}$ & $(29.4, -0.11, 5.90)$\SI{}{\km\per\second} & \cite{OHare:2019qxc}\\
         $\vec{v}_0$         & $(0, 0, 220)$\SI{}{\km\per\second} & \cite{mccabe_astrophysical_2010} \\
         $\vec{v}_\varoast$  & $(11.10, 12.24, 7.25)$\SI{}{\km\per\second} & \cite{OHare:2018trr} \\
         $v_\text{esc}$& \SI{544}{\km\per\second} & \cite{smith_rave_2007} \\\hline\hline
    \end{tabular}
    \label{tab:shmparams}
\end{table}

Recent astrophysical observations indicate that the local DM halo is more complex than is implied by a Maxwell-Boltzmann distribution, as evidenced by kinematically and spatially distinct stellar populations, which likely arose from mergers and accretion locally in the galaxy. 
Simulations indicate that such events may lead to similar substructures in the DM phase space distribution \cite{necib_inferred_2019,necib_under_2019}.
Proposed structures range from cold components like co-rotating DM disks~\cite{read_dark_2009} to hot components like in-falling extragalactic DM, near $\vec{v}_\text{esc}$~\cite{freese_direct_2001}. 

Recent observations from the \Gaia\ survey provide evidence of such substructures that may be in the local halo. 
These observations are often enhanced with additional information from cross-matched observations of the SDSS, which together form the \SSDSGaia\ catalog. 
The substructure classifications used in~\cite{necib_under_2019} are adopted here, based on whether the structures are spatially or kinematically mixed with the SHM.
This classification includes three categories:
(1) Relaxed halo: spatially and kinematically mixed, (2) Debris flows: spatially mixed but kinematically distinct, and (3) Streams: distinct in both space and velocity.
Stellar substructures of all three types have been observed in the \Gaia\ data, and simulations indicate that the relaxed halo and debris flow stellar populations likely act as tracers for similar DM structures.
Stellar streams were also found to trace DM streams, though the correspondence is less strong due to spatial differences between both populations~\cite{necib_under_2019}.
Furthermore, there may be differences in the relative abundance of stars and DM in these substructures, as stars are more tightly bound towards the center of a galaxy than DM is.
This difference may render DM more readily accreted than stars, and the ratio of accreted stellar to DM mass can vary significantly~\cite{necib_under_2019}. 

\subsubsection{Debris flows and streams}
\label{subsubsec:substruc}
This analysis considers various DM velocity distribution functions (VDFs) that may arise due to halo substructures.
These substructures are motivated by observed stellar structures.
In the case of debris flows, the stellar populations provide strong evidence of a similar DM population.
For streams, the correlation between stellar and DM populations is weaker~\cite{necib_under_2019}; observed stellar streams motivate the kinematics of similar DM substructures, but the true properties of the underlying DM streams are less certain.

Because of these uncertainties, results are presented for wide ranges in the overall contribution of each substructure.

For DM streams, galactic-frame VDFs are modeled following the prescription used in~\cite{OHare:2019qxc} as,
\begin{gather}
\begin{aligned}
    f^\text{gal}_{\text{sub}}&(\vec{v}) =  N_\text{sub} \\ &\times\exp\left[-\Big(\vec{v}-\langle\vec{v}_{\text{sub}}\rangle\Big)^T\frac{\bm{\sigma}_\text{sub}^{-2}}{2}\Big(\vec{v}-\langle\vec{v}_{\text{sub}}\rangle\Big)\right] \\
    &\times\Theta(v_\text{esc}-|\vec{v}|)
    \label{eq:streamvdf}
\end{aligned}
\end{gather}
where $N_\text{sub}$ is a normalization constant for the given substructure, $\langle\vec{v}_\text{sub}\rangle$ is the mean velocity of DM particles in the stream or debris flow, and
$\bm{\sigma}_{\text{sub}}$ is its dispersion tensor. Since \DEAP\ is not sensitive to direction, only the total spread in the DM speed and the fraction of the particles' velocity that is parallel to the Earth's velocity affect potential signals.
For simplicity, $\bm{\sigma}_{\text{sub}}$ is therefore taken to be diagonal.

The total VDF for all DM in the halo is given by,
\begin{equation}
    f^\text{gal}_\chi(\vec{v}) = (1-\eta_\text{sub})\cdot f^\text{gal}_\text{SHM}(\vec{v}) + \eta_\text{sub}\cdot f^\text{gal}_\text{sub}(\vec{v})
    \label{eq:totalvdf}
\end{equation}
where \SubstructureFraction\ is the fraction of DM that is in the substructure. Eq. \eqref{eq:totalvdf} ensures that the \textit{total} local DM density $\rho_\chi$ remains fixed, as it is independent of any substructure in phase space distribution.

A number of stellar streams have been identified in astronomical measurements; in these cases, these observed streams are used to motivate values for $\langle\vec{v}_\text{sub}\rangle$ and $\bm{\sigma}_\text{sub}$. 
Streams considered are discussed below and listed in Tab.~\ref{tab:vdfs}.
The effects of each stream on the WIMP exclusion curves were studied, and streams with similar effects were grouped together. 

The following  substructures are considered:
\begin{enumerate}[(a)]
    \item \textbf{\GaiaSausage}, also known as the \Gaia\ Enceladus~\cite{10.1093/mnras/sty982} or GRASP \cite{Bozorgnia:2019mjk}, likely results from a merger event with a massive dwarf galaxy (\SI{\sim5e10}{\msun}) at a redshift of $z\lesssim3$~\cite{myeong_sausage_2018, necib_inferred_2019}. The VDF is best fit with a bimodal distribution comprising two Gaussian distributions. Ref.~\cite{Necib:2019zka} showed that this structure appears to extend all the way into the galactic plane, suggesting that it should be correlated with local substructure in the dark sector.  The single-Gaussian parameterization in~\cite{OHare:2019qxc} and the directly inferred (bimodal) VDF presented in~\cite{necib_inferred_2019} are considered.
    \item \textbf{S1 stream} 
    is a counter-rotating (retrograde) stellar stream, likely from a progenitor with a stellar mass around \SIrange{e6}{e7}{\msun}, potentially related to the $\omega$ Centauri globular cluster~\cite{Myeong:2017skt} or the Fornax dwarf spheroidal galaxy~\cite{10.1093/mnras/sty1403}. Evidence suggests that the stellar component passes through the local neighborhood, potentially indicating a significant local DM component, as well~\cite{OHare:2018trr}. The VDF used for this stream is described in~\cite{OHare:2019qxc}, and it is represented by G2 in Tab.~\ref{tab:vdfs}.
    
    \item \textbf{Nyx} is a co-rotating (prograde) stellar stream, lagging behind the Sun by \SI{\sim80}{\km\per\second}, and appearing to intersect the solar neighborhood~\cite{Necib:2019zbk}. This stellar stream may indicate a similar DM stream, which is described using the parametrization in~\cite{Necib:2019zka}. Nyx is represented by G6 in Tab.~\ref{tab:vdfs}.
    
    \item \textbf{Helmi Stream} is a significant stellar stream identified in the solar neighborhood in several galactic surveys, and it may indicate a similar substructure in the local DM halo~\cite{Helmi:2008eq}. Simulations favor an origin from a merger event with a \SI{\sim e8}{\msun} dwarf galaxy around \SIrange{5}{8}{Gyr} ago~\cite{koppelman_characterization_2019}. The parametrization used for these studies is from~\cite{Necib:2019zka}, which identifies ``Group I'' with the Helmi Stream. In Tab.~\ref{tab:vdfs}, it is represented by G5.
    
    \item \textbf{Koppelman 1} and \textbf{Koppelman 2} are a pair of stellar streams identified in the solar neighborhood, first identified in~\cite{Koppelman_2018}. They appear to be from relatively recent accretion events. In this study, these streams are parameterized using the values given in ~\cite{Necib:2019zka}, where they are referred to as ``Group II'' and ``Group III''. In Tab.~\ref{tab:vdfs}, Koppelman 1 and Koppelman 2 are represented by G1 and G2, respectively.
\end{enumerate}

\subsubsection{In-falling clumps}

A generic model of ``in-falling clumps'' (ICs) is considered, describing extra-galactic DM accreted into the Milky Way, not described by the observed stellar streams.
Such ICs have been proposed by a number of authors ~\cite{sikivie_secondary_1997,freese_direct_2001,stiff_signatures_2001,gelmini_weakly_2001,green_potential_2001,Lisanti:2011as,Kuhlen:2012fz,baushev_extragalactic_2013}, and may arise from past merger events or from intergalactic DM continually falling into the Milky Way, as motivated by models of hierarchical galaxy formation.
To investigate the effects of ICs,  galactic-frame VDFs are modeled using Eq.~\eqref{eq:streamvdf}, with mean velocity $\langle\vec{v}_\text{sub}\rangle = \left(\langle v_r\rangle,\langle v_\theta\rangle, \langle v_\phi\rangle\right)$ and dispersion tensor $\bm{\sigma}_\text{sub} = \text{diag}(\sigma_{rr}, \sigma_{\theta\theta}, \sigma_{\phi\phi})$ given in galactocentric spherical coordinates, with $r$ pointing towards the center of the galaxy, $\theta$ describing the zenith angle, and $\phi$ oriented with the disk rotation, and components given by,

\begin{gather}
    \begin{aligned}\label{eq:icvdf}
    \langle v_\phi\rangle &= |\vec{v}|\cos\alpha \\
    \langle v_r\rangle &= \langle v_\theta\rangle = \frac{1}{\sqrt{2}}|\vec{v}|\sin\alpha \\
    \sigma_{\phi\phi} &= \sigma_{||}\cos\alpha + \sigma_\perp\sin\alpha \\ 
    \sigma_{rr} &= \sigma_{\theta\theta} =\frac{1}{\sqrt{2}}\left( \sigma_{||}\sin\alpha - \sigma_\perp\cos\alpha\right) \\
    \sigma_{ij} &= 0 \text{ , if } i \neq\ j
    \end{aligned}
\end{gather}
where $|\vec{v}|$ is the magnitude of the mean velocity vector, $\alpha$ is the angle between this vector and the motion of the Earth, $\sigma_{||}$ is the dispersion of the IC parallel to the Earth's velocity, and $\sigma_\perp$ is the dispersion in the perpendicular directions. 

To reduce the number of parameters, components of $\langle\vec{v}_\text{sub}\rangle$ and $\bm{\sigma}_\text{sub}$ that are perpendicular to the Earth's motion are set equal to each other. 
While this equality is not guaranteed for a generic VDF, a temporally-averaged direct detection experiment insensitive to recoil direction is only sensitive to a DM particle's speed and the fraction of the velocity parallel to the Earth's motion.
Changing the velocity division between the $r$- and $\phi$-directions has a negligible impact on the resulting exclusion curves.

ICs were considered with all 27 combinations of $\alpha\in\{0,\pi/2,\pi\}$, $|\vec{v}|\in\{200, 300, 400\}$ \SI{}{\kilo\meter\per\second}, and $\sigma_{||}\in\{10, 30, 50\}$ \SI{}{\kilo\meter\per\second}, with $\sigma_\perp$ fixed to \SI{50}{\kilo\meter\per\second}, chosen as a typical value from the range of streams considered in Tab.~\ref{tab:vdfs}. 
The three values chosen for $\alpha$ correspond to ICs that enter the galaxy in prograde, perpendicular, or retrograde directions.
Total speeds below $v_\text{esc}$ were considered, with the understanding that some energy would be lost to dynamical friction as the ICs accreted into the galaxy.
Values for $\sigma_{||}$ were investigated less than or equal to $\sigma_{\perp}$, under the assumption that as streams become elongated, their phase space density is approximately conserved following Liouville's theorem, and the streams become correspondingly colder.

\subsubsection{VDF Groupings}

\begin{table*}[htb]
    \centering
    \setcellgapes{2pt}\makegapedcells \renewcommand\theadfont{\normalsize\bfseries}
    \caption{Summary of substructures considered in this study. The mean velocity vector for each Galactic-frame VDF is given as $\langle\vec{v}_\text{sub}\rangle=\left(\langle v_r\rangle, \langle v_\theta\rangle, \langle v_\phi\rangle\right)$, and the dispersion tensor is defined as $\bm{\sigma}_\text{sub}=\text{diag}\left(\sigma_{rr}, \sigma_{\theta\theta}, \sigma_{\phi\phi}\right)$. The mass fraction of the local DM in each substructure is \SubstructureFraction; the total DM density is kept constant at $\rho_\chi=\SI{0.3}{\GeV\per\square\c\per\cubic\cm}$. Streams and in-falling clumps (ICs) are arranged in groups based on similar effects on exclusion curves; these groups are denoted by G$N$, where $N$=\numrange{1}{6}. Two models of the \GaiaSausage\ are considered, as described by~\cite{necib_inferred_2019} and~\cite{OHare:2019qxc}. For \GaiaSausage\ (Necib~\etal), the numerical VDF provided in~\cite{necib_inferred_2019} was used, and the values describing $\langle\vec{v}_\text{sub}\rangle$ and $\bm{\sigma}_\text{sub}$ are quoted for comparison. For all other substructures, the listed parameters were used as input to Eq.~\eqref{eq:streamvdf}. Values are given as described in the references. Where numbers were given with quoted uncertainties, the central value was used; where ranges were provided, the midpoint of the range was considered. ICs are given as ``IC ($\alpha$, $|\vec{v}|$)'', and are defined as described in Eq.~\eqref{eq:icvdf}, with $\sigma_{||}=\SI{30}{\km\per\second}$ and $\sigma_\perp=\SI{50}{\km\per\second}$. Substructures chosen to represent each group are marked with $^*$. To model Koppelman 1 and Helmi VDFs, the central value of the dispersion components was used.}
    
    \begin{tabular}{clcc | ccc | ccc | c }\hline
      & \multirow{2}{*}{Substructure} & \multirow{2}{*}{Type} & \multirow{2}{*}{Ref.} &$v_r$ & $v_\theta$ & $v_\phi$ & $|\sigma_{rr}|$ & $|\sigma_{\theta\theta}|$ & $|\sigma_{\phi\phi}|$ & \multirow{2}{*}{\SubstructureFraction} \\ 
      & & & & \multicolumn{3}{c|}{[\SI{}{\km\per\second}]} & \multicolumn{3}{c|}{[\SI{}{\km\per\second}]} & \\\hline\hline

     & \GaiaSausage\ (Necib~\etal)  & Debris flow & \cite{necib_inferred_2019} & \SI{\pm 147}{}$^{+7.2}_{-6.4}$ & \SI{-2.8}{}$^{+1.5}_{-1.6}$ & \SI{27.9}{}$^{+2.8}_{-2.9}$ & \SI{113.6}{}$^{+3.1}_{-3.0}$ & \SI{65.2}{}$^{+1.1}_{-1.2}$ & \SI{61.9}{}$^{+2.6}_{-2.9}$ & \numrange{0}{0.70}  \\\hline
     & \GaiaSausage\ (O'Hare~\etal) & Debris flow & \cite{OHare:2019qxc}       & \SI{-8.2}{} & \SI{0.99}{} & \SI{25.7}{} & \SI{158.9}{} & \SI{80.9}{} & \SI{61.5}{} & \numrange{0}{0.70}   \\\hline

     \multirow{2}{*}{\rotatebox[origin=c]{90}{\scriptsize G1}} & Koppelman 1$^*$  & Stream & \cite{Necib:2019zka} & \SI{-169}{} & \SI{-59}{} & \SI{-375}{} & \numrange{11}{37} & \numrange{3}{16} & \numrange{6}{28} & \numrange{0}{0.30} \\
     & IC $\left(\pi, \SI{400}{\km\per\second}\right)$ & IC & --- & 0 & 0 & \SI{-400}{} & 35.4 & 35.4 & 30 & \numrange{0}{0.30}\\\hline

     \multirow{3}{*}{\rotatebox[origin=c]{90}{\scriptsize G2}} & S1$^*$            & Stream & \cite{OHare:2019qxc} & \SI{-29.6}{} & \SI{-72.8}{} & \SI{-297.4}{} & \SI{82.6}{} & \SI{58.5}{} & \SI{26.9}{} & \numrange{0}{0.30} \\
     & Koppelman 2   & Stream & \cite{Necib:2019zka} & \SI{213}{} & \SI{161}{} & \SI{-226}{} & \SI{52}{} & \SI{18}{} & \SI{29}{} & \numrange{0}{0.30} \\
     & IC $\left(\pi, \SI{300}{\km\per\second}\right)$ & IC & --- & 0 & 0 & \SI{-300}{} & 35.4 & 35.4 & 30 & \numrange{0}{0.30} \\\hline
     
     \multirow{2}{*}{\rotatebox[origin=c]{90}{\scriptsize G3}} & IC $\left(\pi, \SI{200}{\km\per\second}\right)^*$ & IC & --- & 0 & 0 & 200 & 35.4 & 35.4 & 30 & \numrange{0}{0.30} \\
     & IC $\left(\frac{\pi}{2}, \SI{400}{\km\per\second}\right)$ & IC & --- & 282.8 & 282.8 & 0 & 21.2 & 21.2 & 50 & \numrange{0}{0.30} \\\hline
     
     \multirow{1}{*}{\rotatebox[origin=c]{90}{\scriptsize G4}} & IC $\left(\frac{\pi}{2}, \SI{300}{\km\per\second}\right)^*$ & IC & --- & 212.1 & 212.1 & 0 & 21.2 & 21.2 & 50 & \numrange{0}{0.30} \\\hline
     
     \multirow{2}{*}{\rotatebox[origin=c]{90}{\scriptsize G5}} & Helmi$^*$      & Stream & \cite{Necib:2019zka} & \SI{29}{} & \SI{-287}{} & \SI{141}{} & \numrange{37}{83} & \numrange{6}{21} & \numrange{4}{15} & \numrange{0}{0.30} \\
     & IC $\left(\frac{\pi}{2}, \SI{200}{\km\per\second}\right)$ & IC & --- & 141.4 & 141.4 & 0 & 21.2 & 21.2 & 50 & \\\hline
     
     \multirow{5}{*}{\rotatebox[origin=c]{90}{\scriptsize G6}} & Nyx$^*$           & Stream & \cite{Necib:2019zbk} & \SI{156.8}{}$^{+2.1}_{-2.2}$ & \SI{-1.4}{}$^{+3.1}_{-3.0}$ & \SI{141.0}{}$^{+2.5}_{-2.6}$ & \SI{46.9}{}$^{+1.7}_{-1.6}$ & \SI{70.9}{}$^{+2.4}_{-2.2}$ & \SI{52.5}{}$^{+1.8}_{-1.8}$ & \numrange{0}{0.30}    \\
     & IC ($0$, \SI{400}{\km\per\second}) & IC & --- & 0 & 0 & \SI{-400}{} & 35.4 & 35.4 & 30 & \numrange{0}{0.30} \\
     & IC ($0$, \SI{300}{\km\per\second}) & IC & --- & 0 & 0 & \SI{-300}{} & 35.4 & 35.4 & 30 & \numrange{0}{0.30} \\
     & IC ($0$, \SI{200}{\km\per\second}) & IC & --- & 0 & 0 & \SI{-200}{} & 35.4 & 35.4 & 30 & \numrange{0}{0.30} \\\hline\hline
    \end{tabular}
    \label{tab:vdfs}
\end{table*}

To reduce the number of exclusion curves drawn, substructures with similar VDFs were arranged into groups.
To determine the optimal grouping, sample exclusion curves were drawn for the $\mathcal{O}_1$ and $\mathcal{O}_5$ operators (discussed in Section~\ref{subsec:eft}) assuming each substructure is present at the maximum relative density considered.
Curves for ICs and streams naturally formed groups with similar behavior to each other.
One representative VDF from each group was then selected to be used for the full analysis, presented in Sec. \ref{sec:results}.
These groups are summarized in Tab.~\ref{tab:vdfs}.

For the ICs, varying $\sigma_{||}$ in the range considered had very little effect on the exclusion curves.
Therefore, only ICs with $\sigma_{||}=\SI{30}{\km\per\second}$ are further considered.
Similarly, all prograde ICs had nearly identical results; this is due to the fact that their mean velocity in the Earth's frame gave DM particles too little kinetic energy to produce a signal in the WIMP-search region of interest.

In~\cite{OHare:2019qxc}, it is argued that streams may contribute up to \SI{20}{\percent} of the local stellar population.
Given the weaker correlation between stars and DM in streams, it is possible that the observed stellar substructures under- or over-represent the underlying DM populations, and so possible relative densities \SubstructureFraction\ are evaluated in the range \SIrange{0}{30}{\percent}.

Two proposed VDFs describing the \GaiaSausage are considered: Necib~\etal~\cite{necib_inferred_2019}, for which the numerical VDF was obtained from \cite{gitLNecib}, and O'Hare~\etal~\cite{OHare:2019qxc} (first described in~\cite{evans_shm:_2018}), which provides the parameters quoted in Tab. \ref{tab:vdfs}.
These two descriptions significantly differ in their suggested values for \SubstructureFraction.
In~\cite{necib_inferred_2019}, the authors arrive at \SubstructureFraction$=42^{+26}_{-22}$\SI{}{\percent},  Ref.~\cite{evans_shm:_2018} proposes that \SubstructureFraction$=20\pm10\SI{}{\percent}$, and a best-fit value of \SubstructureFraction$=\SI{61}{\percent}$ is obtained in~\cite{OHare:2019qxc}, comparing the relative weights of the Sausage and ``round halo'' components.
To cover the full range of possibilities, \SubstructureFraction\ is considered in the range \SIrange{0}{70}{\percent}.

All of the VDFs under consideration are shown in Fig.~\ref{fig:allvdf}, for comparison.
VDFs with similar impacts on exclusion curves are given the same color and grouped together as in Tab.~\ref{tab:vdfs}.

\subsection{\label{subsec:eft}Effective operators}
The NREO approach (or nonrelativistic effective field theory, NREFT) is a method of parametrizing the set of possible contact interactions governing DM-nucleon interactions that may arise from a full theory of dark matter~\cite{Fan:2010gt,fitzpatrick_effective_2013,Fitzpatrick:2012ib,Anand:2013yka,Dent:2015zpa}. 
Chiral Effective Field Theory (ChEFT) provides an alternative approach, accounting for one- and two-body currents, described in~\cite{hoferichter_analysis_2016,Hoferichter:2018acd}. One-body currents described by ChEFT can be mapped to NREFT operators and are therefore included in this framework; two-body currents are not considered in the present analysis.
The NREFT Hamiltonian can include terms that couple coherently to the nucleus and the DM, as well as to the DM spin $\vec S_\chi$, the nucleon spin $\vec S_{N}$, the exchanged momentum $\vec q$, and the component \vvperp\ of the relative velocity $\vec{v}_\text{rel}$ that is orthogonal to $\vec q$:
\begin{equation}
    \vvperp \cdot \vec{q} \equiv 0
\end{equation}
such that
\begin{equation}
    \vvperp = \vec{v}_\text{rel} - \frac{\vec{q}}{2 m_N}
\end{equation}
where $\vec{v}_\text{rel}$ comes from Eq.~\eqref{eq:totalvdf} boosted to the Earth frame and $m_N$ is the nucleon mass.

By convention, the set of scalar combinations of these vector operators is labeled as $\CO_{i}$. 
The dimensionful couplings to each operator are denoted as $c_i^\tau$, where $\tau\in\{0,1\}$ represents the isoscalar and isovector components, respectively, $i$ denotes the NREO index, and subscripts $\chi$ and $N$ refer to operators acting on DM and nucleons, respectively.

$\CO_1$ is the standard spin-independent (SI) interaction, where DM coherently scatters with all nucleons; $\CO_4 = \vec{S}_\chi \cdot \vec{S}_N$ is the spin-dependent operator, which gives cross sections proportional to the total nuclear spin $J$, which is 0 for \arforty.
Operators that can lead to \NRs\ with \arforty\ are~\cite{fitzpatrick_effective_2013},
\begin{gather}
\begin{aligned}
\CO_1 &= 1_\chi 1_N  \\
\CO_3 &= i \vec{S}_N \cdot ({ \frac{\vec{q}} {m_N}} \times \vvperp) \\
\CO_5 &= i \vec{S}_\chi \cdot ({\frac{\vec{q}} {m_N}} \times \vvperp) \\
\CO_8 &= \vec{S}_\chi \cdot \vvperp   \\
\CO_{11} &= i \vec{S}_\chi \cdot {\frac{\vec{q}} {m_N}} \\
\CO_{12} &=  \vvperp \cdot (\vec{S}_\chi \times \vec{S}_N) \\
\CO_{15} &= - \left(\vec{S_\chi} \cdot \frac{\vec{q}}{m_N}\right)\left[\left(\vec{S}_N\times\vvperp\right)\cdot \frac{\vec{q}}{m_N}\right]
\label{eq:operators}
\end{aligned}
\end{gather}

Operators that depend on $\vec{S}_N$ can still lead to scattering on \arforty.
For example, $\CO_3$ is sensitive to spin-orbit coupling, rather than nuclear spin. 

Following the prescription used in~\cite{fitzpatrick_effective_2013,Catena:2015uha,Hoferichter:2018acd}, the operators in Eq.~\eqref{eq:operators} give rise to the DM-nucleus scattering cross section, which is factored into DM response functions $R^{\tau\tau^\prime}_k\left( \vperpsquare,q^2\right)$ and nuclear response functions $W^{\tau\tau^\prime}_k\left(q^2\right)$, with $|\vec{S}_\chi|=1/2$,
\begin{gather}
\begin{aligned}
    \frac{d\sigma}{dE_R}=& \frac{2m_T}{v^2}\sum_{\tau,\tau^\prime}\left\{R_M^{\tau\tau^\prime}\left(\vperpsquare,q^2\right)W_M^{\tau\tau^\prime}\left(q^2\right) \right. \\
    &+ \frac{q^2}{m_N^2} \left[R^{\tau\tau^\prime}_{\Phi^{\prime\prime}}\left(\vperpsquare,q^2\right)W_{\Phi^{\prime\prime}}^{\tau\tau^\prime}\left(q^2\right)\right. \\
    & \qquad+ \left.\left.  R^{\tau\tau^\prime}_{\Phi^{\prime\prime}M}\left(\vperpsquare,q^2\right)W_{\Phi^{\prime\prime}M}^{\tau\tau^\prime}\left(q^2\right) \right] \right\},
\end{aligned}
\label{eq:dsigmadE}
\end{gather}
only including terms that are nonzero for \arforty. 
The $W^{\tau\tau^\prime}_k\left(q^2\right)$ terms are computed in~\cite{Catena:2015uha} using nuclear shell model techniques~\cite{Hoferichter:2018acd,Catena:2015uha} for each interaction with \arforty.
They are given as best-fit polynomials. 
Subscripts $k = M, \Phi^{\prime\prime}, M\Phi^{\prime\prime}$ represent different one-body multipole operators in the nuclear matrix element. $M$ is the standard spin-independent nuclear response, which describes the nucleon density inside the nucleus. It coherently sums over all nucleons and closely resembles the Helm form factor. At zero-momentum transfer, $\Phi''$ is related to the angular momentum and spin ($\vec{L}\cdot\vec{S}$) of nucleons. It favors heavier elements with large, not fully occupied, spin-partner angular-momentum orbitals. It can be of the same order as the $M$ response for heavier elements.

The accuracy of the nuclear shell model used to compute $W^{\tau\tau^\prime}_k(q^2)$ and the implications for nuclear structure factor calculations are discussed in~\cite{Hoferichter:2018acd}. In these studies, it is shown that the hierarchy of states and low-energy observables such as the charge radius are well-reproduced for several nuclei, including \arforty. The authors of~\cite{Hoferichter:2018acd} conclude that nuclear shell model uncertainties are not expected to have a significant effect on the ground states involved in \WIMP-nucleus scattering. Furthermore, comparisons between $W_M^{00}(q^2)$ and the experimentally-motivated Helm form factor suggested in~\cite{lewin_review_1996} show that both form factor calculations agree to within \SI{0.5}{\percent} in the energy range of interest to the present study. Uncertainties in the nuclear response functions derived from the nuclear shell model are therefore assumed to be negligible in this analysis.

The $R^{\tau\tau^\prime}_k$ terms are calculated in~\cite{fitzpatrick_effective_2013,Catena:2015uha}, and depend on the coupling strengths of the operators in Eq.~\eqref{eq:operators}. Keeping only the terms that contribute for \arforty, these response functions are given in Eq.~\eqref{eq:dmresponse}.
\begin{widetext}
\begin{gather}
\begin{aligned}
R^{\tau\tau^\prime}_M\left(\vperpsquare, q^2\right) &= c_1^\tau c_1^{\tau^\prime} + \frac{j_\chi(j_\chi+1)}{3}\left(\frac{q^2}{m_N^2}\vperpsquare c_5^\tau c_5^{\tau^\prime} + \vperpsquare c_8^\tau c_8^{\tau^\prime} + \frac{q^2}{m_N^2}c_{11}^{\tau}c_{11}^{\tau^\prime}\right)
\\
R^{\tau\tau^\prime}_{\Phi^{\prime\prime}}\left(\vperpsquare ,q^2\right) &= \frac{q^2}{4m_N^2}c_3^\tau c_3^{\tau^\prime} + \frac{j_\chi(j_\chi+1)}{12} \left(c_{12}^\tau - \frac{q^2}{m_N^2}c_{15}^\tau\right)
\left(c_{12}^{\tau^\prime} - \frac{q^2}{m_N^2}c_{15}^{\tau^\prime}\right) 
\\
R^{\tau\tau^\prime}_{\Phi^{\prime\prime}M}\left(\vperpsquare ,q^2\right) &= c_3^\tau c_1^{\tau^\prime} + \frac{j_\chi(j_\chi+1)}{3}\left(c_{12}^\tau - \frac{q^2}{m_N^2}c_{15}^\tau\right)c_{11}^{\tau^\prime}.
\label{eq:dmresponse}
\end{aligned}
\end{gather}
\end{widetext}

The present analysis places limits on couplings to $\CO_1, \CO_3, \CO_5, \CO_8$, and $\CO_{11}$, considering only one coupling at a time. 
$\CO_{12}$ and $\CO_{15}$ are not included, since in EFTs they always arise in combination with other operators that will dominate the scattering process \cite{Dent:2015zpa}. 
Furthermore, for certain models leading to $\CO_{10}$, $\CO_{11}$, and $\CO_{12}$, loop contributions to the neutron electric dipole moment lead to constraints that are orders of magnitude stronger than those from direct detection experiments~\cite{Drees:2019qzi}.

The standard isoscalar SI interaction $\CO_1$, discussed in~\cite{Ajaj:2019imk}, relies only on the coupling constant $c_1^0$, and the response function $R^{00}_MW^{00}_M\left(q^2\right)$ is comparable to the square of the Helm form factor~\cite{lewin_review_1996}.

Results of the present study are presented in terms of an effective DM-proton cross section
\begin{eqnarray}\label{eqn:effsig}
\sigma_p  \equiv \frac{(c^p_i \: \mu_p)^2}{\pi},
\label{eq:sigeff}
\end{eqnarray}
where $c^p_i \equiv (c^0_i + c^1_i)/2$ is the effective DM-proton coupling and $\mu_p$ is the DM-proton reduced mass. Note that although Eq.~\eqref{eq:sigeff} does not explicitly depend on the coupling to neutrons $c^n_i \equiv (c^0_i - c^1_i)/2$, limits placed on $\sigma_p$ implicitly depend on the value of $c^n_i$. Unless explicitly specified, $c^n_i = c^p_i$, corresponding to isoscalar ($c^1_i = 0$) couplings. 

Even though Eq.~\eqref{eq:sigeff} gives the standard SI DM-nucleon cross section for $\CO_1$, it does not necessarily correspond to a physical cross section for all possible interactions. This relation is used because it allows for a direct comparison between operators and experiments, and it gives a one-to-one correspondence to  couplings in an effective Hamiltonian. 

\subsubsection{\label{subsec:isospin}Isospin violation}
As mentioned above, the NREFT framework allows for general isospin couplings: $c_i^0$ corresponds to the \emph{isoscalar} coupling (IS), while $c_i^1$ would be an \emph{isovector} coupling (IV). Varying the ratio of $c^1_i/c^0_i$---or equivalently $c^n_i/c^p_i$---can lead to different DM-nucleus couplings to different elements and isotopes. 
Isospin-violating DM has been considered as a way to reconcile disparate experimental results, and arise, for example, in non-\WIMP\ SUSY DM models~\cite{Feng:2011vu}. 
While typical direct detection results only report limits on IS couplings, other couplings are also possible.

The present analysis considers IS ($c_i^n = c_i^p$) and IV ($c_i^n = - c_i^p$) scenarios, as well as \textit{xenonphobic} (XP) interactions $c_i^n/c_i^p = -0.7$. Since the strongest SI limits on DM-nucleon scattering are currently from xenon-based experiments, it is worth examining the parts of parameter space that would not yield a strong signal in xenon.
The scenario was proposed \cite{Feng:2013vod,Gresham:2013mua} as a way to explain potential direct detection anomalies in light of the strong bounds from XENON100 \cite{Aprile:2012nq} and LUX \cite{Akerib:2012ak}. A number of theoretical models have been built that can result in such isospin-violating interactions. These include interference between two distinct portals to the dark sector \cite{Belanger:2013tla}, new colored mediators \cite{Hamaguchi:2014pja}, string theory-motivated $Z'$ portal scenarios \cite{Martin-Lozano:2015vva}, a two-higgs doublet portal model \cite{Drozd:2015gda}, and the coupling ratio resulting from breaking of GUT-scale gauge groups \cite{Li:2019sty}.

In~\cite{Yaguna:2019llp}, it was shown that, as such a ratio minimizes the DM-xenon scattering cross section, it brings the sensitivity of \DEAP\ beyond that of \Xenon\ for \WIMPMassSymbol$>$\SI{130}{\GeV\per\square\c}, as long as isospin-dependent $W^{\tau\tau^\prime}_k$ effects are small.

\subsubsection{Photon-mediated interactions}
\label{subsec:longrange}
The NREFT formalism does not directly cover the case of light mediators, where the momentum dependence of the propagator becomes important and long-range forces can lead to a signal enhancement. 
These interactions can nonetheless be parametrized in terms of NREOs. 
In addition to the operators listed above, anapole, electric/magnetic dipole, and millicharge interactions are considered, taking their nonrelativistic limits as in \cite{DelNobile:2013sia,DelNobile:2014eta,Buch:2019aiw}. 

The anapole interaction can be written:
\begin{eqnarray}
\mathcal{O_A} =  e \, c_\mathcal{A} \sum_{N=n,p} \left( Q_N \mathcal{O}_8 + g_N \mathcal{O}_9\right),
\label{eq:oanapole}
\end{eqnarray}
where $e$ is the charge of the electron, $Q_n=0$ and $Q_p=1$, $c_\mathcal{A}$ is the Wilson coefficient of the effective anapole interaction, and $g_n = -3.83$ and $g_p = 5.59$ are the nucleon $g$-factors.

In the presence of a magnetic dipole moment $\mu_\chi$, the relevant effective operator is:
\begin{eqnarray}
\mathcal{O_{MD}} &=&  2 \, e \, {\mu_\chi}  \sum_{N=n,p} \left[ Q_N \, m_N \mathcal{O}_1 + 4 \, Q_N \frac{m_\chi m_N^2}{q^2} \mathcal{O}_5 \right. \nonumber  \\  &&+  \left. 2\, g_N \, m_\chi \left (\mathcal{O}_4 - \frac{m_N^2}{q^2} \mathcal{O}_6 \right) \right] ,
\label{eq:omdipole}
\end{eqnarray}
and the electric dipole moment, $d_\chi$, gives rise to the non-relativistic electric operator, $\mathcal{O_{ED}}$: 

\begin{eqnarray}
\label{eq:electricD}
\mathcal{O_{ED}} =  2 \, e \, d_\chi \, \frac{\mathcal{O}_{11}^{(p)}}{q^2}.
\label{eq:oedipole}
\end{eqnarray}

Finally, the millicharged interaction leads to a standard Rutherford interaction:
\begin{eqnarray}
\mathcal{O_M} = e^2 \, \epsilon_\chi  \, \frac{\mathcal{O}_1^{(p)}}{q^2},
\label{eq:omillicharge}
\end{eqnarray}
where $\epsilon_\chi$ is the DM electric charge relative to the electron charge.
The operators with the superscript $(p)$ in Eqs.~\eqref{eq:oedipole} and~\eqref{eq:omillicharge} only couple to protons.
In each case, the total cross section is computed using DM and nuclear response functions for each NREO.

\section{\label{sec:analysis} Analysis and methods}

VDFs are numerically constructed in the laboratory frame using a Monte Carlo method. 
For each VDF, 2\E{7} particles are generated in the galactic rest frame using Eq.~\eqref{eq:totalvdf}, and then boosted to the laboratory frame. 

Differential cross sections are calculated using Eq.~\eqref{eq:dsigmadE}, considering one coupling at a time, with nuclear response functions determined by~\cite{Catena:2015uha}.
Calculations are validated by computing recoil energy spectra assuming the SHM and comparing the results to those from the \wimpy~\cite{gitBJK} and \chiraleftdm~\cite{Hoferichter:2018acd} public codes, where possible.

The integral in Eq.~\eqref{eq:diffscat} is numerically computed, using the \NR\ acceptance determined for the latest \WIMP\ search by \DEAP~\cite{Ajaj:2019imk}. This calculation determines the expected number of DM-induced \NR\ signals expected for a given set of models.

This study uses the same data set and analyses reported in~\cite{Ajaj:2019imk}, with no \WIMP-like events remaining after all event selection cuts.
These null results provide constraints on the DM-nucleon coupling constants for the $\CO_1$, $\CO_3$, $\CO_5$, $\CO_8$, and $\CO_{11}$ operators, which are interpreted as effective cross sections using Eq~\eqref{eqn:effsig}.

For photon-mediated interactions (Eqs.~(\ref{eq:oanapole}-\ref{eq:omillicharge})), limits are placed on the anapole coupling constant $c_\mathcal{A}$, the magnetic dipole moment $\mu_\chi$, the electric dipole moment $d_\chi$, or the relative electric charge $\epsilon_\chi$.

Upper limits are reported at the \SI{90}{\percent} confidence level (\CL); systematic uncertainties from the detector response model, signal acceptance, and exposure are propagated into the upper limit following the prescription by Cousins and Highland~\cite{cousins_incorporating_1992}, and are detailed in~\cite{Ajaj:2019imk}.

\section{\label{sec:results}Results and discussion}

This section reports the key findings of this analysis.  Sec.~\ref{subsec:spectra} shows the predicted effects of the different NREOs and VDFs on \NR\ spectra. Secs.~\ref{subsec:nreoconstraints}, \ref{subsec:IVconstraints} and \ref{subsec:vdfconstraints} respectively show the constraints obtained assuming different NREFT operators, isospin violation scenarios, and VDFs. 
Sec.~\ref{subsec:allISconstraints} illustrates the interplay between NREOs and nonstandard VDFs, and Sec.~\ref{subsec:allthemodels} adds isospin violation. Finally, Sec.~\ref{subsec:specificinteractionconstraints} shows limits on DM with a magnetic dipole, anapole interaction, electric dipole and fractional charge.

\subsection{Recoil energy spectra for different interactions and VDFs}
\label{subsec:spectra}

The expected \NR\ energy spectra can be calculated using Eq.~\eqref{eq:diffscat}. 
They depend on the underlying VDF as well as the DM and nuclear response functions, as written in Eq.~\eqref{eq:dsigmadE}.

\begin{figure}[htb]
    \centering
    \includegraphics[width=\linewidth]{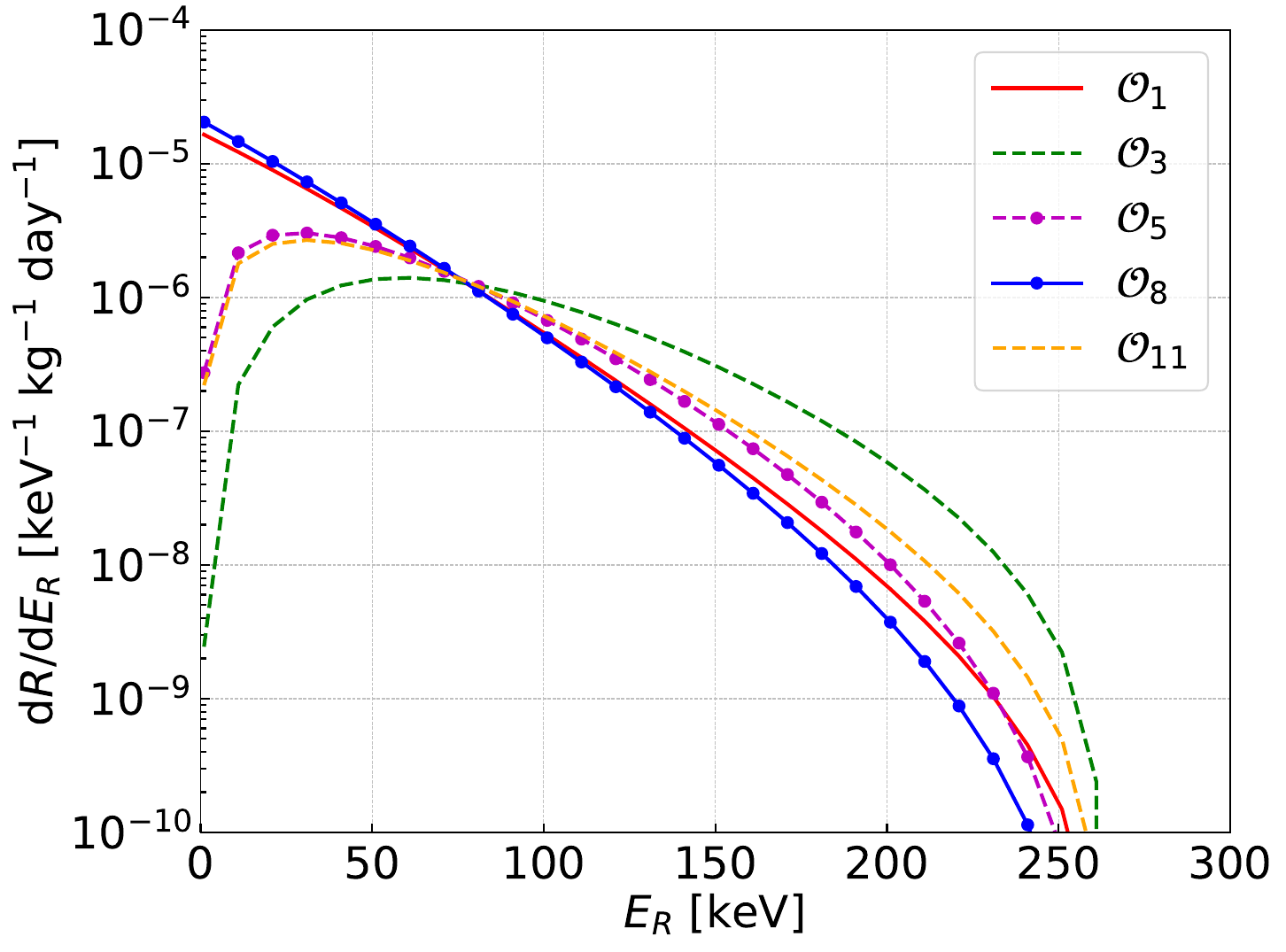} 
    \caption{Recoil spectra for \WIMPMassSymbol=\SI{100}{\GeV\per\square\c} and SHM, using the following cross sections: 
    $\CO_1$ (\SI{3.7e-45}{\square\cm}, red), 
    $\CO_3$ (\SI{3.1e-38}{\square\cm}, green), 
    $\CO_5$ (\SI{2.9e-36}{\square\cm}, purple), 
    $\CO_8$ (\SI{1.9e-38}{\square\cm}, blue), and
    $\CO_{11}$ (\SI{2.3e-42}{\square\cm}, orange).  }
    \label{fig:rate_effope}
\end{figure}

Fig.~\ref{fig:rate_effope} shows the recoil energy spectra for \WIMPs\ with \WIMPMassSymbol=\WIMPMassHundredGeV\ that interact with nucleons via different NREOs, assuming the SHM.
These spectra are normalized to cross sections that predict a similar number of events in the energy region of interest.
Operators that introduce a factor of $q^2$ to the DM response function ($\CO_5$, $\CO_{11}$) are suppressed at low recoil energies, exhibiting a peak around \SI{25}{\keV}, while for $\CO_3$ ($\sim q^4$) the peak is around 50 keV.
Operators that add a factor of \vperpsquare\ ($\CO_5$ and $\CO_8$) have qualitatively little effect on the recoil spectra, though the spectra drop off slightly faster, due to the fact that \vperp\ suppresses backscattering.

\begin{figure}[htb]
    \centering
    \includegraphics[width=\linewidth]{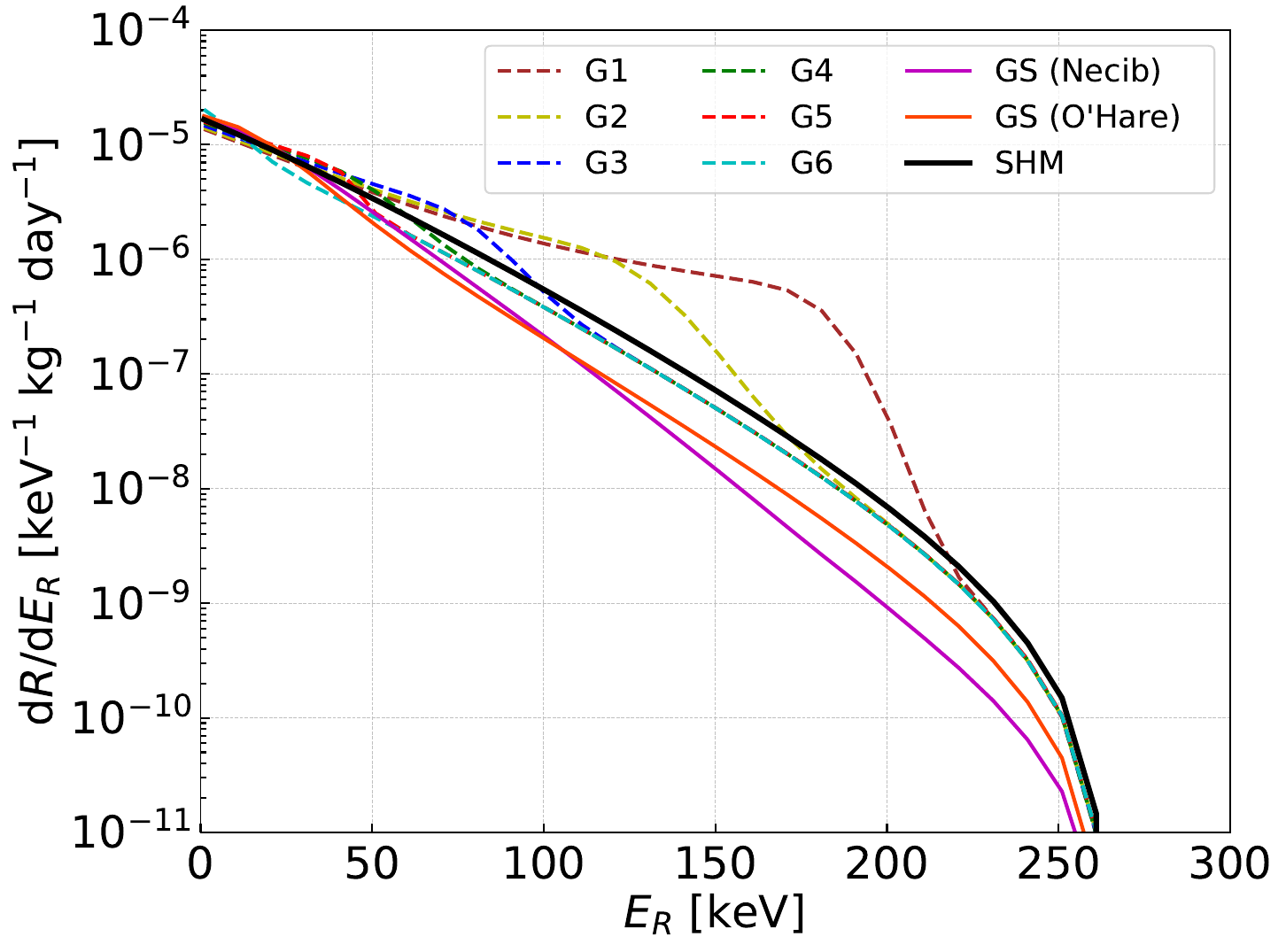} 
    \caption{Recoil spectra for \WIMPMassSymbol=\SI{100}{\GeV\per\square\c} and $\CO_{1}$ with different substructures at maximum \SubstructureFraction, using a cross section of \SI{3.7e-45}{\square\cm}. Curves labeled ``GS'' correspond to the two \GaiaSausage\ models.}
    \label{fig:rate_vdf_O1}
\end{figure}

Effects of substructures on the $\CO_1$ recoil spectrum are illustrated in Fig.~\ref{fig:rate_vdf_O1}, where each substructure has been taken at its maximum \SubstructureFraction.
Spectra from slow substructures (the \GaiaSausage, G4, G5, and G6) decrease faster than predicted by the SHM, while those resulting from fast substructures (G1, G2, and G3) are flattened by a knee around \SIrange{75}{175}{\keV}.
While these distortions affect the expected rate of recoils in the energy region of interest, the spectra in this range are similar. 

\begin{figure}[htb]
    \centering
    \includegraphics[width=\linewidth]{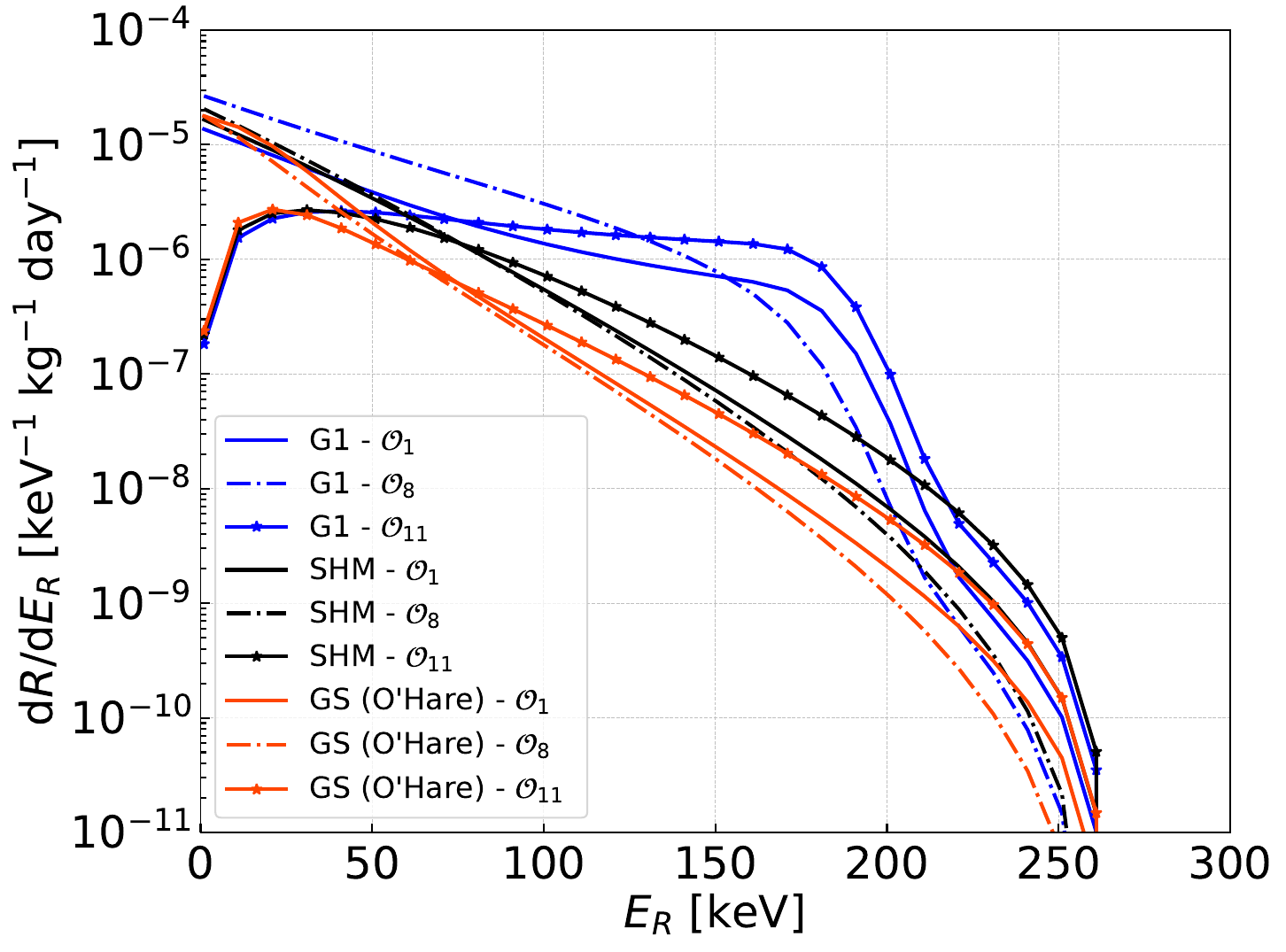} 
    \caption{Recoil spectra for \WIMPMassSymbol=\SI{100}{\GeV\per\square\c} with VDFs from G1 (blue), SHM (black), and \GaiaSausage~\cite{OHare:2019qxc} (``GS (O'Hare)'', red), with maximum \SubstructureFraction, and the following cross sections: 
    $\CO_1$ (\SI{3.7e-45}{\square\cm}, solid), 
    $\CO_8$ (\SI{1.9e-38}{\square\cm}, dash-dot), and
    $\CO_{11}$ (\SI{2.3e-42}{\square\cm}, solid-star). }
    \label{fig:rateVDFs_O1O8O11}
\end{figure}

Fig.~\ref{fig:rateVDFs_O1O8O11} shows the effect that substructures may have for $\CO_1$, $\CO_8$, and $\CO_{11}$.
These operators were selected, for their respective scaling factors of 1, \vperpsquare, and $q^2$.
The effects of the \GaiaSausage\ from~\cite{OHare:2019qxc} and G1 streams are compared to the spectra derived from the pure SHM, assuming the maximum considered value of \SubstructureFraction.
These substructures were chosen to span the range of low- and high-speed VDFs.

The spectra from $\CO_8$ and $\CO_{11}$ are more strongly affected by these substructures than $\CO_1$.
For $\CO_{11}$, the \GaiaSausage\ causes the recoil spectrum to decrease nearly exponentially, at a faster rate than the SHM alone predicts, while G1 renders it near flat in the range of \SIrange{25}{175}{\keV}.
This shape is a result of the higher momentum transfers accessible by the fast components of G1; the cross section for such interactions is increased by the $q^2$-enhancement of this operator.
However, the energy region of interest used in this study extends to \SI{100}{\keV}, below much of this enhancement. 

Stronger effects are observed for $\CO_8$, for which the nuclear scattering cross section scales with $v_\perp^2$. 
In this case, fast DM particles in G1 have an enhanced nuclear scattering cross section, even when the momentum transfer is relatively small.
This behavior leads to enhanced cross sections across all accessible energy scales.
Similarly, substructures like the \GaiaSausage\ that decrease the amount of fast DM suppress the recoil spectrum.

\begin{figure}[htb]
    \centering
    \includegraphics[width=\linewidth]{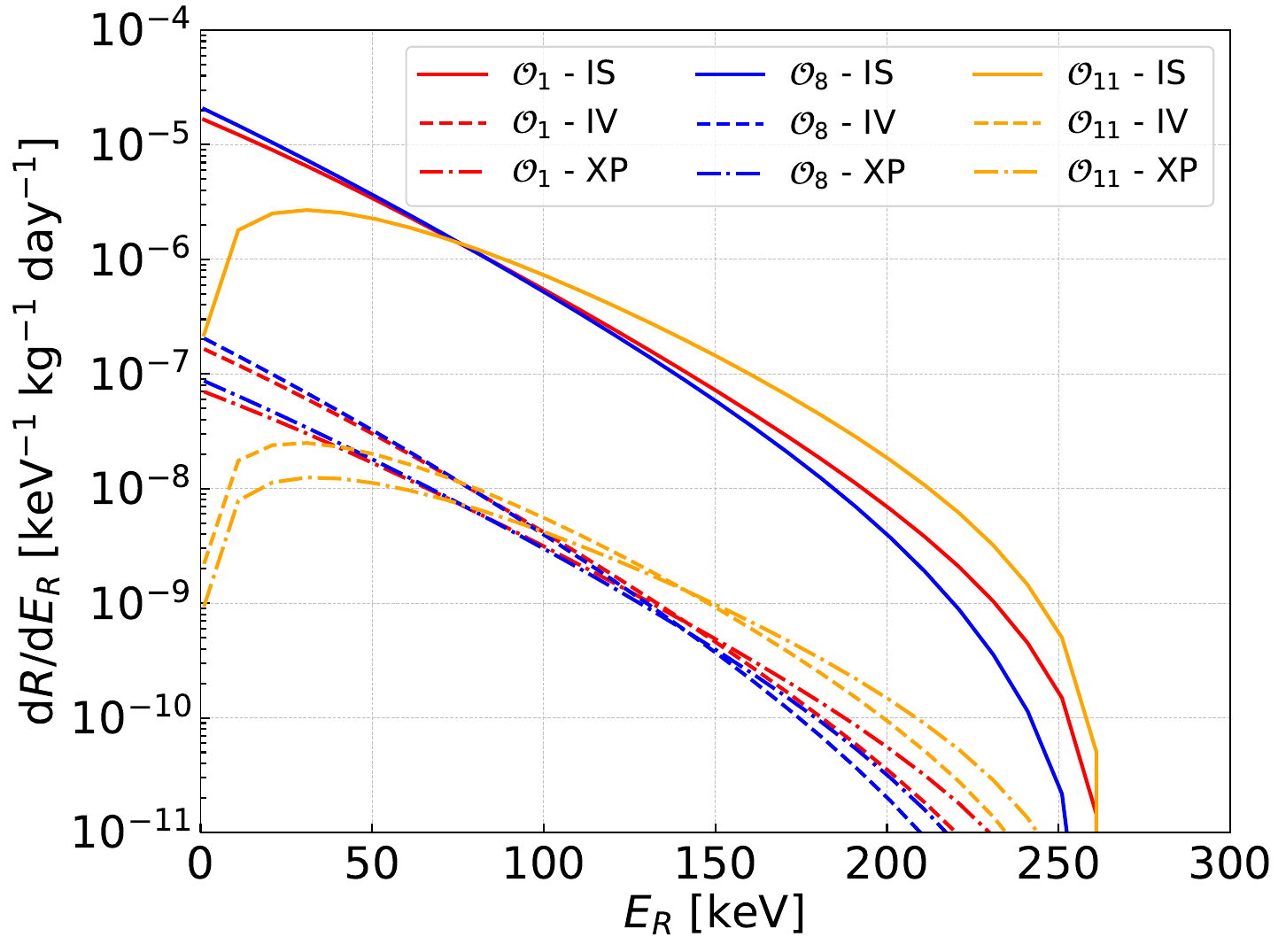} 
    \caption{Recoil spectra for \WIMPMassSymbol=\SI{100}{\GeV\per\square\c} and SHM with IS, IV, and XP couplings, and the following cross sections: 
    $\mathcal{O}_{1}$ (\SI{3.7e-45}{\square\cm}, red),
    $\mathcal{O}_{8}$ (\SI{1.9e-38}{\square\cm}, blue), and
    $\mathcal{O}_{11}$ (\SI{2.3e-42}{\square\cm}, orange).}
    \label{fig:rate_vdf_isop}
\end{figure}

Fig.~\ref{fig:rate_vdf_isop} shows the recoil energy spectra for DM-nucleon couplings via the same three operators, assuming the SHM, for IS, IV, and XP isospin scenarios.
Similar behavior is observed for all three operators.
IS interactions have the strongest nuclear couplings, due to the coherent $A^2$ enhancement (where $A$ is the atomic mass number), while interference between protons and neutrons suppress IV and XP interactions.
These interactions all have slightly different shapes, governed by their corresponding nuclear response function $W_k^{\tau\tau^\prime}(q^2)$ in Eq.~\eqref{eq:dsigmadE}.
These functions are defined for IS and IV components, as well as their cross terms, which appear in XP interactions.
The IV term decreases the most quickly with recoil energy, while the cross terms are relatively flat. 
As a result, the IV energy spectrum decreases the fastest, while the XP spectrum (the only one including the cross terms) decreases the slowest.

\subsection{Constraints on effective operators, with the Standard Halo Model}
\label{subsec:nreoconstraints}

\begin{figure}[htb]
    \centering
    \includegraphics[width=\linewidth]{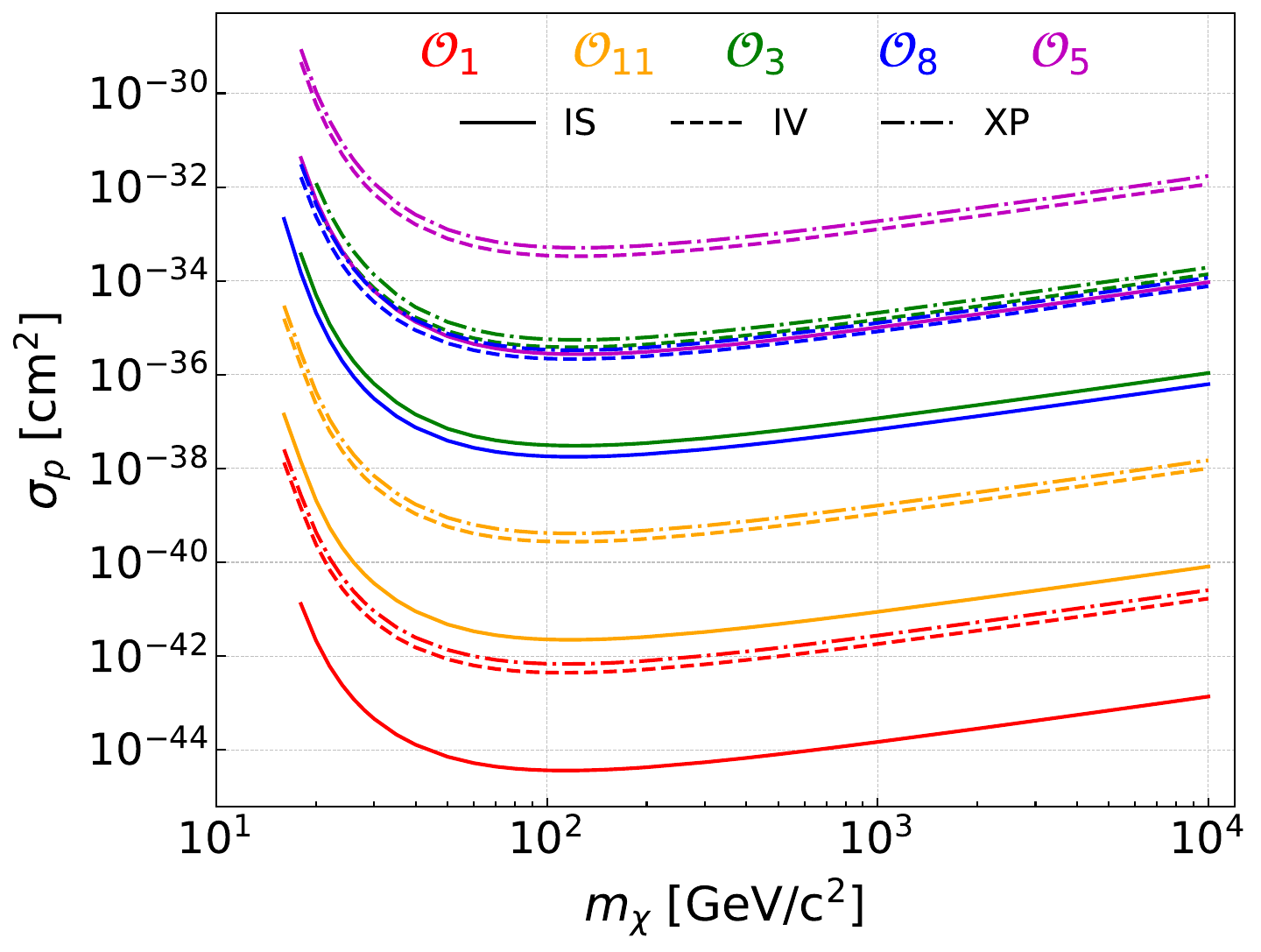} 
    \caption{Upper limits (\SI{90}{\percent} C.L.) on DM-nucleon scattering cross sections with the SHM and operators: $\mathcal{O}_1$ (red), $\mathcal{O}_{11}$ (orange), $\mathcal{O}_3$ (green), $\mathcal{O}_8$ (blue) and $\mathcal{O}_5$ (purple).
    IS interactions (solid lines) always set the strongest constraints. Isospin-violating scenarios (IV: dash lines, and XP: dash-dot lines) are also shown.}
    \label{fig:limeverything}
\end{figure}

Exclusion curves for the  NREOs considered here are presented in Fig.~\ref{fig:limeverything} for the SHM VDF, as a function of the DM mass $m_\chi$ and the effective DM-proton cross section $\sigma_p$ defined in Eq. \eqref{eq:sigeff}. 

Operators $\CO_1$, $\CO_5$, $\CO_8$, and $\CO_{11}$ depend on the $M$ response function, and $\CO_3$ depends on $\Phi^{\prime\prime}$.
As can be seen in Fig.~\ref{fig:limeverything}, interactions governed by $\CO_5$, $\CO_8$ and $\CO_{11}$ are suppressed relative to $\CO_1$, despite using the same nuclear response function. 
This suppression is due to the additional factor of $(q/m_N)^2\sim10^{-3}$--$10^{-2}$ in $\CO_{11}$ and the factor of $\vperpsquare \sim 10^{-6}$ in $\CO_8$; while both factors suppress $\CO_5$. 

The operator $\CO_3$ is proportional to $(q/m_N)^4$, while $\CO_{11}$ goes as $(q/m_N)^2$. $\CO_3$ is described by the $\Phi^{\prime\prime}$ multipole operator (discussed in Eqs.~\eqref{eq:dsigmadE} and~\eqref{eq:dmresponse}), while $\CO_{11}$ is described by $M$.  Since the former operator is related to spin-orbit coupling, it couples to the two unpaired neutrons and proton holes in \arforty, rather than to all 40 nucleons. This leads to a suppression of $\sim 10^2$ in addition to the extra $q^2$ suppression.

These results can be compared to those reported by \DSf\ in~\cite{darkside-50_collaboration_effective_2020}, where similar behavior was observed.
The study in~\cite{darkside-50_collaboration_effective_2020} also explores the effects of light mediators in these interactions.
The analysis presented by \DSf\ adopts a different convention for interpreting effective coupling constants as DM-nucleon cross sections than is used in the present study (see Eq.~\eqref{eq:sigeff})---namely, \DSf\ provides IS cross sections normalized to reference values for $q$ and \vperp\ at $q_\text{ref}=$\SI{100}{\MeV\per\c} and $v_\text{ref}=$\SI{220}{\km\per\second}, respectively.
Recasting the IS constraints shown in Fig.~\ref{fig:limeverything} using these conventions shows that the present constraints are stronger, as expected from the increased exposure used for the present search.

\subsection{Effects of isospin violation on constraints}
\label{subsec:IVconstraints}

\begin{figure}[htb]
    \centering
    \includegraphics[width=\linewidth]{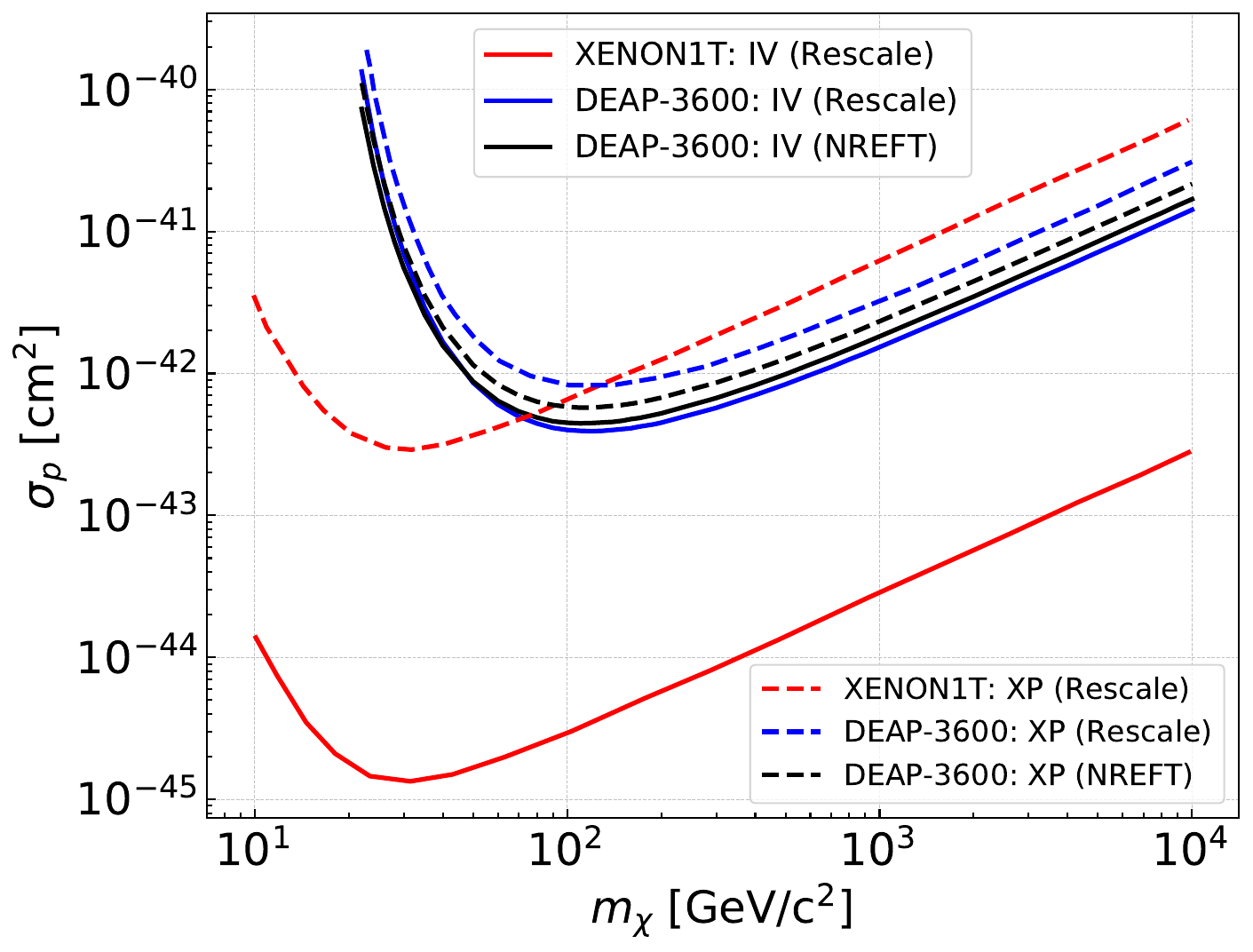}
    \caption{Constraints on the $\CO_1$ interaction from \Xenon~\cite{aprile_dark_2018} and \DEAP~\cite{Ajaj:2019imk}, for IV (isovector; solid) and XP (xenonphobic; dashed) scenarios. Limits labeled ``Rescale'' were obtained following the method used in~\cite{Yaguna:2019llp} (shown in Eq. \ref{eq:sigmapprime}), while those labeled ``NREFT'' used the present approach. 
    }
    \label{fig:XPDM}
\end{figure}

The effects of IV and XP isospin scenarios on constraints on the DM-nucleon cross section resulting from $\CO_1$ with the pure SHM are illustrated in Fig.~\ref{fig:XPDM}. 
As seen in Fig.~\ref{fig:rate_vdf_isop}, the recoil energy spectra for the IV and XP scenarios differ, due to the different nuclear response functions produced by IS and IV couplings, as well as their cross terms.

Isospin violation in \LAr\ and \LXe\ targets was explored in~\cite{Yaguna:2019llp}, where DM-nucleon cross sections were rescaled to various isospin scenarios using previously reported constraints on the isoscalar DM-nucleon cross section $\sigma_N^\text{IS} = (c^0_i\mu_{N})^2/\pi$.
These are related by the ``rescaling method'':
\begin{eqnarray}
\label{eq:sigmapprime}
\sigma_p'=\sigma_N^\text{IS}\frac{\sum_j \eta_j \mu_{A_j}^2A_j^2}{\sum_j \eta_j\mu_{A_j}^2\left[Z+(A_j-Z)c^n_i/c^p_i\ratio\right]^2} ,
\end{eqnarray}
where $\eta_j$ is the relative abundance of isotope $j$ with mass number $A_j$ and $Z$ protons; and $\mu_{A_j}$ is the reduced mass of the DM-nucleus system.

To further explore these effects, Fig.~\ref{fig:XPDM} shows limits for IV and XP scenarios with the ``rescale'' method in~\cite{Yaguna:2019llp} and with the NREFT framework.
The results of rescaling limits from \Xenon~\cite{aprile_dark_2018} and \DEAP~\cite{Ajaj:2019imk} are shown for the IV and XP scenarios, consistent with values obtained in~\cite{Yaguna:2019llp}.
These rescaled constraints from \DEAP\ are compared with constraints obtained from the NREFT framework where $\sigma_p'$ is defined in Eq.~\eqref{eq:sigmapprime} and NR functions $W_k^{\tau \tau'}$ are implemented consistently. As seen in Fig.\ref{fig:XPDM}, this latter framework gives up to \SI{20}{\percent} stronger limits. This difference is due to the fact that the XP recoil energy spectrum is flatter than the IV spectrum, as shown in Fig.~\ref{fig:rate_vdf_isop}.

\subsection{Effects of substructures on constraints}
\label{subsec:vdfconstraints}
\begin{figure}[!htb]
  \centering
  \subfigure[$\CO_1$ interactions]{
    \includegraphics[width=0.97\linewidth]{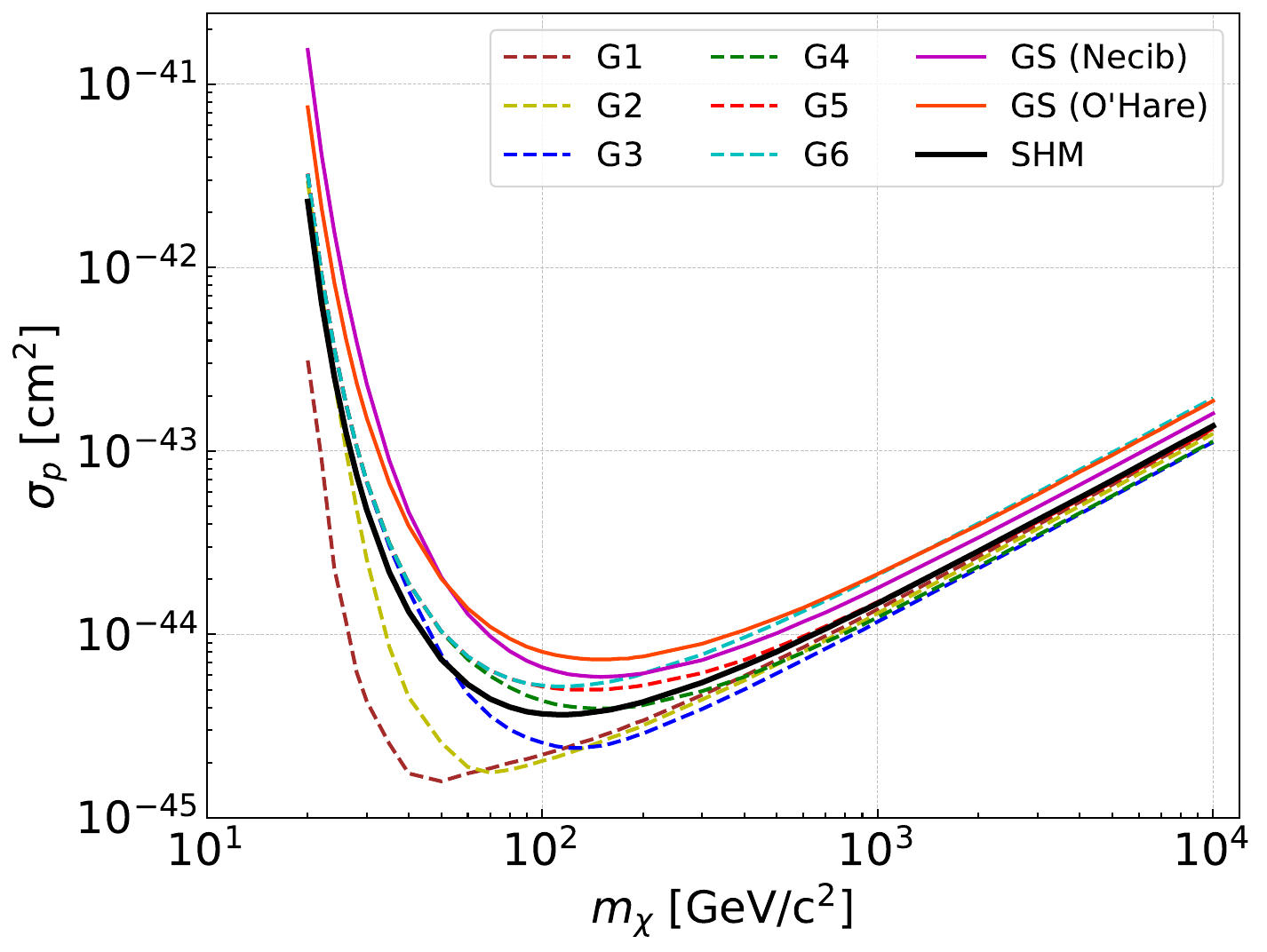} 
    \label{fig:selecvdfO1}
  } \\
  \subfigure[$\CO_5$ interactions]{
    \includegraphics[width=0.97\linewidth]{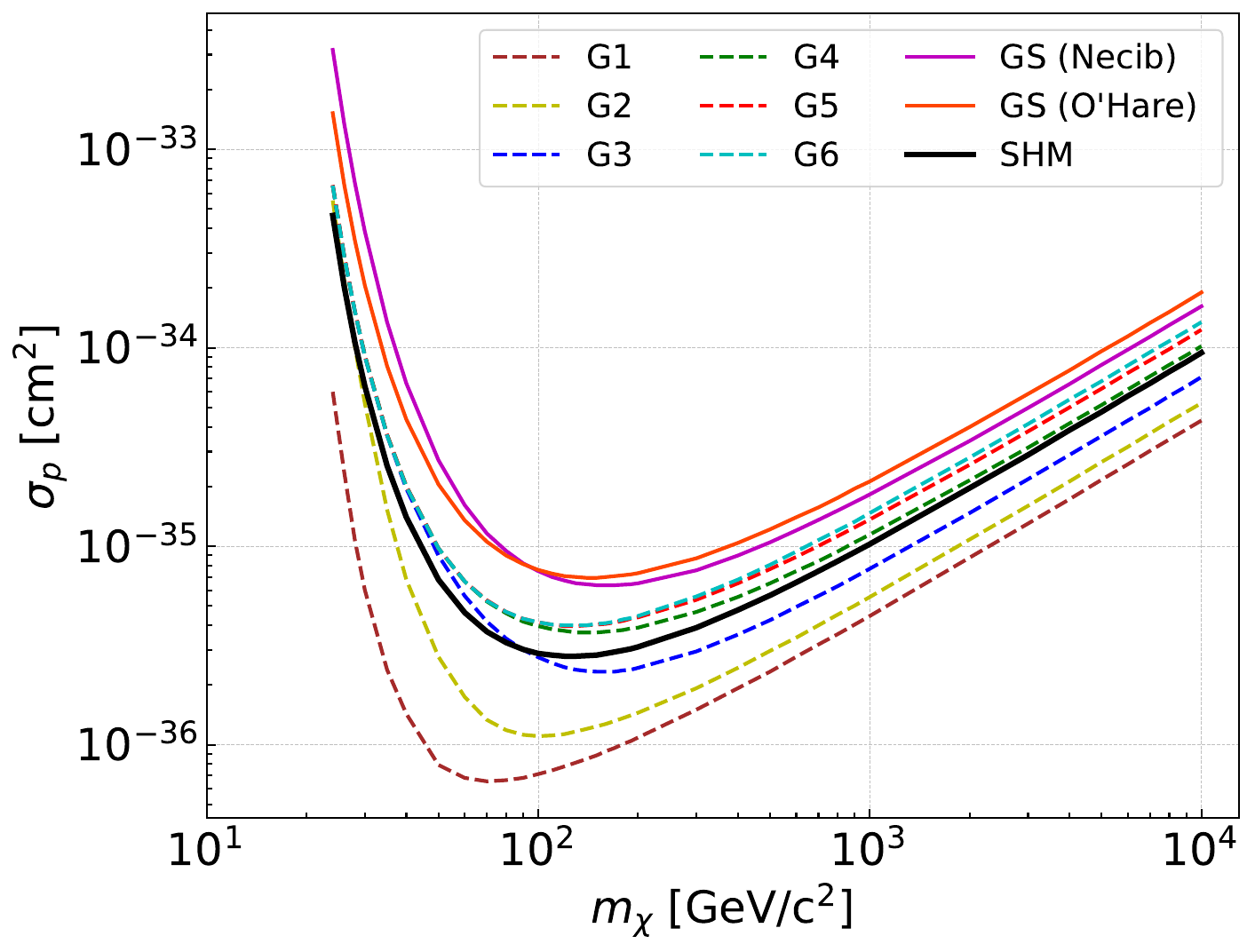}  
    \label{fig:selecvdfO5}
  }
  \caption{Upper limits (\SI{90}{\percent} C.L.) on the effective operators (a) $\CO_1$ and (b) $\CO_5$ for substructures in this study, as defined in Tab.~\ref{tab:vdfs}. Curves labeled ``GS'' correspond to the two \GaiaSausage\ models.
  }
  \label{fig:selecvdfs}
\end{figure}

Fig.~\ref{fig:selecvdfO1} shows the effects of various DM halo substructures on  cross section upper limits for the IS $\CO_1$ interaction, using the maximum values of \SubstructureFraction.

The strongest effects are seen at lower \WIMPMassSymbol, where the lower DM kinetic energy places the  maximum recoil energy closer to the energy threshold.
As such, slow substructures weaken the limits at low \WIMPMassSymbol, while fast ones strengthen them.
These effects diminish at higher \WIMPMassSymbol, where a higher fraction of the DM will have enough kinetic energy to produce visible signals, until they level off at some constant deviation from the limits derived with the SHM.
Once slow particles have enough kinetic energy to reliably produce detectable signals, the effects of increasing their velocity become smaller.
As a result, streams modeled by G1, G2, and G3 lead to stronger limits, while both \GaiaSausage\ models and the streams G4, G5, and G6 result in weaker limits. 

Fig.~\ref{fig:selecvdfO5} illustrates how these limits change when $\CO_5$ is considered, instead. 
Each substructure is again taken at its maximum \SubstructureFraction. 
A similar trend is observed, in which faster substructures lead to stronger limits and slower substructures lead to weaker limits.
However, the effects are much more significant, due to the dependence of $\CO_5$ on \vperpsquare\, with more than an order of magnitude variation seen near  \WIMPMassSymbol $\approx$ \SI{100}{\GeV\per\square\c}. 

These differences persist at higher masses.
For operators that depend on \vperp, enhancing the high-velocity component of the VDF increases the
fraction of candidates with enough kinetic energy to produce a detectable signal, as for $\CO_1$.
These high-velocity DM particles also have enhanced nuclear scattering cross section, yielding stronger constraints. 
Likewise, slower substructures have DM with suppressed interactions and weaker limits.

\subsection{Isoscalar limits in the presence of halo substructures}
\label{subsec:allISconstraints}
\begin{figure*}
  \centering
  \subfigure[\GaiaSausage\ (Necib~\etal)~\cite{necib_inferred_2019}]{
    \includegraphics[width=0.48\linewidth]{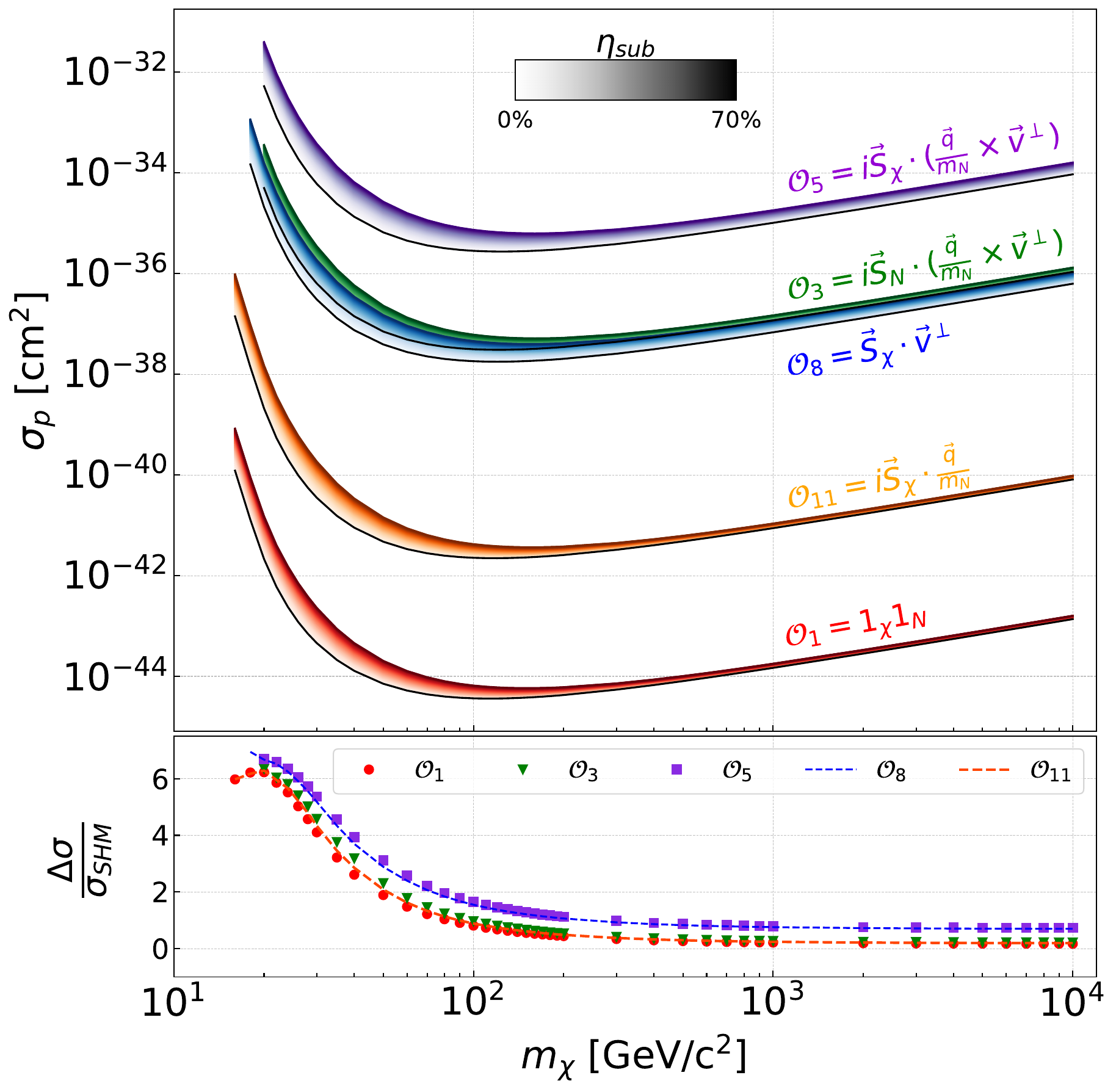}
    \label{fig:limGSLN}
  }
  \subfigure[\GaiaSausage\ (O'Hare~\etal)~\cite{OHare:2019qxc}]{
    \includegraphics[width=0.48\linewidth]{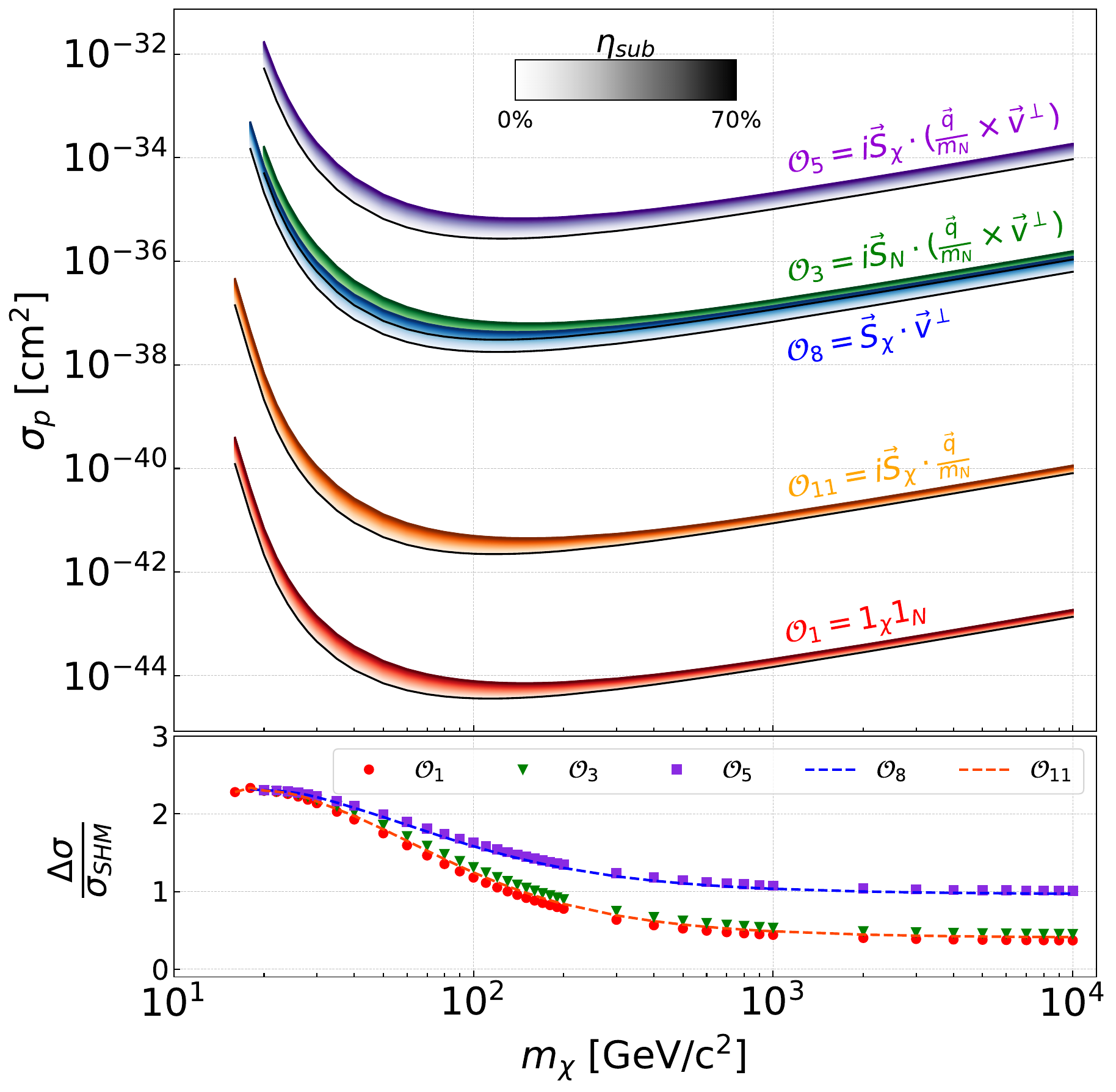}
    \label{fig:limGSOH}
  } \\
  \subfigure[G1 streams]{
    \includegraphics[width=0.48\linewidth]{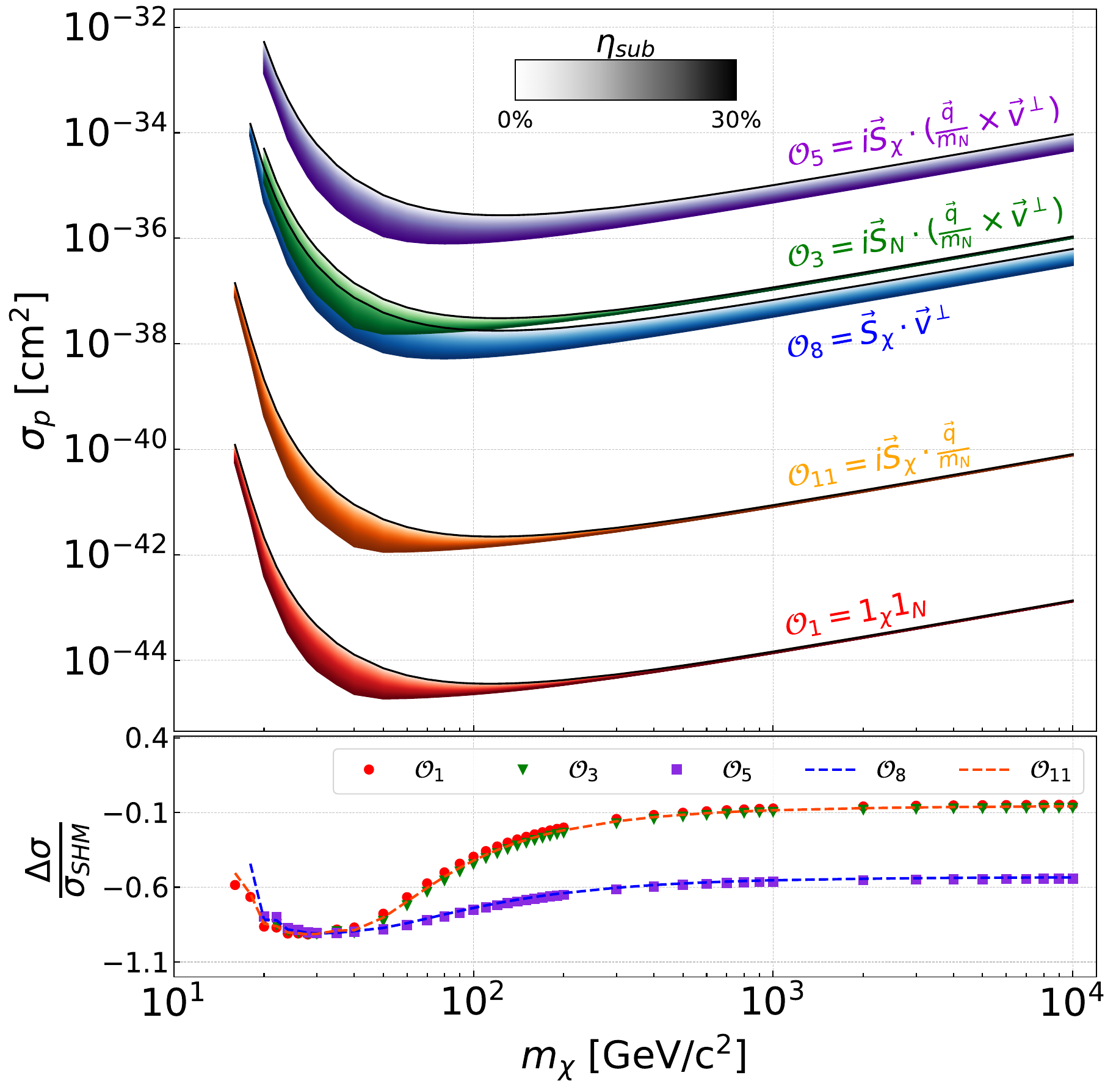}
    \label{fig:limKop}
  }
  \subfigure[G2 streams]{
    \includegraphics[width=0.48\linewidth]{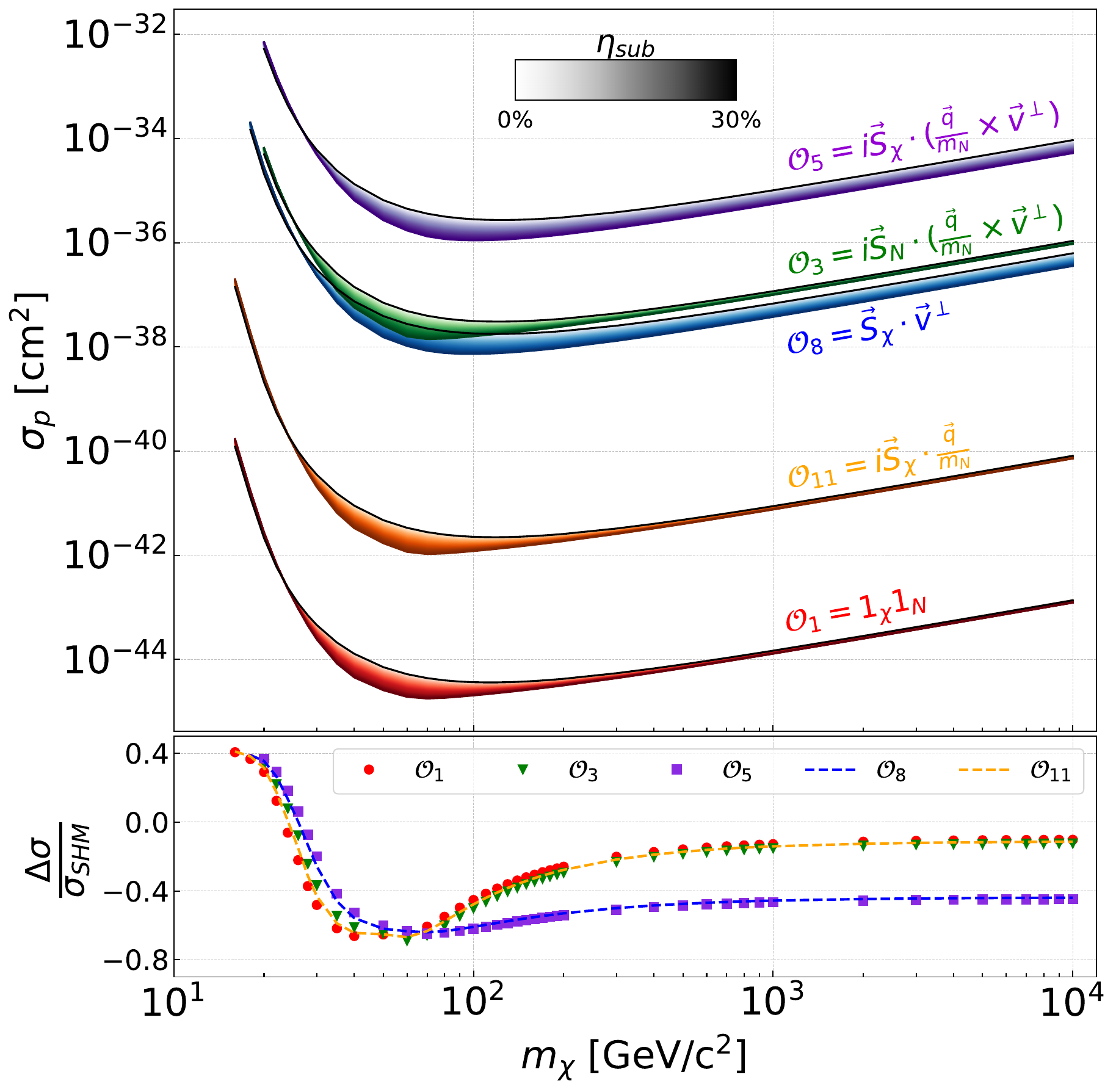}
    \label{fig:limS1}
  }
  \caption{Upper limits (\NinetyPerCentCL) on DM-nucleon scattering cross sections for the $\CO_1$, $\CO_{11}$, $\CO_3$, $\CO_8$, and $\CO_{5}$ effective operators, in the presence of VDFs corresponding to both \GaiaSausage\ models, G1 streams, and G2 streams, with \SubstructureFraction\ of the DM contained in the specified substructure. Beneath each set of exclusion curves is the relative deviation of each operator with the given substructure at its maximum value compared to the SHM and where $\Delta\sigma$ = $\sigma_\text{sub} - \sigma_\text{SHM}$.}
  \label{fig:exclusions1}
\end{figure*}
\begin{figure*}
  \centering
  \subfigure[G3 streams]{
    \includegraphics[width=0.48\linewidth]{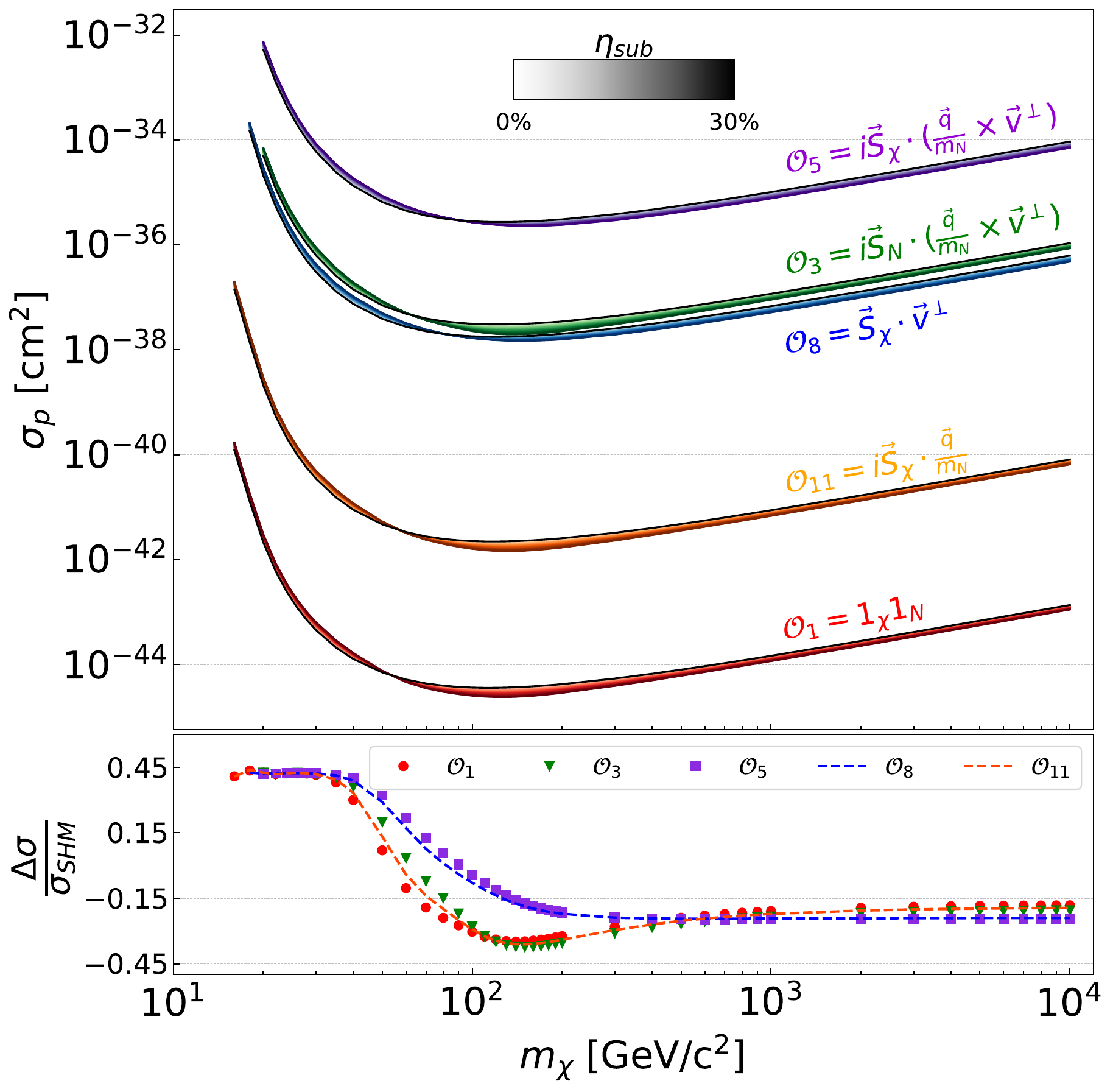}
    \label{fig:limIC2}
  }
  \subfigure[G4 streams]{
    \includegraphics[width=0.48\linewidth]{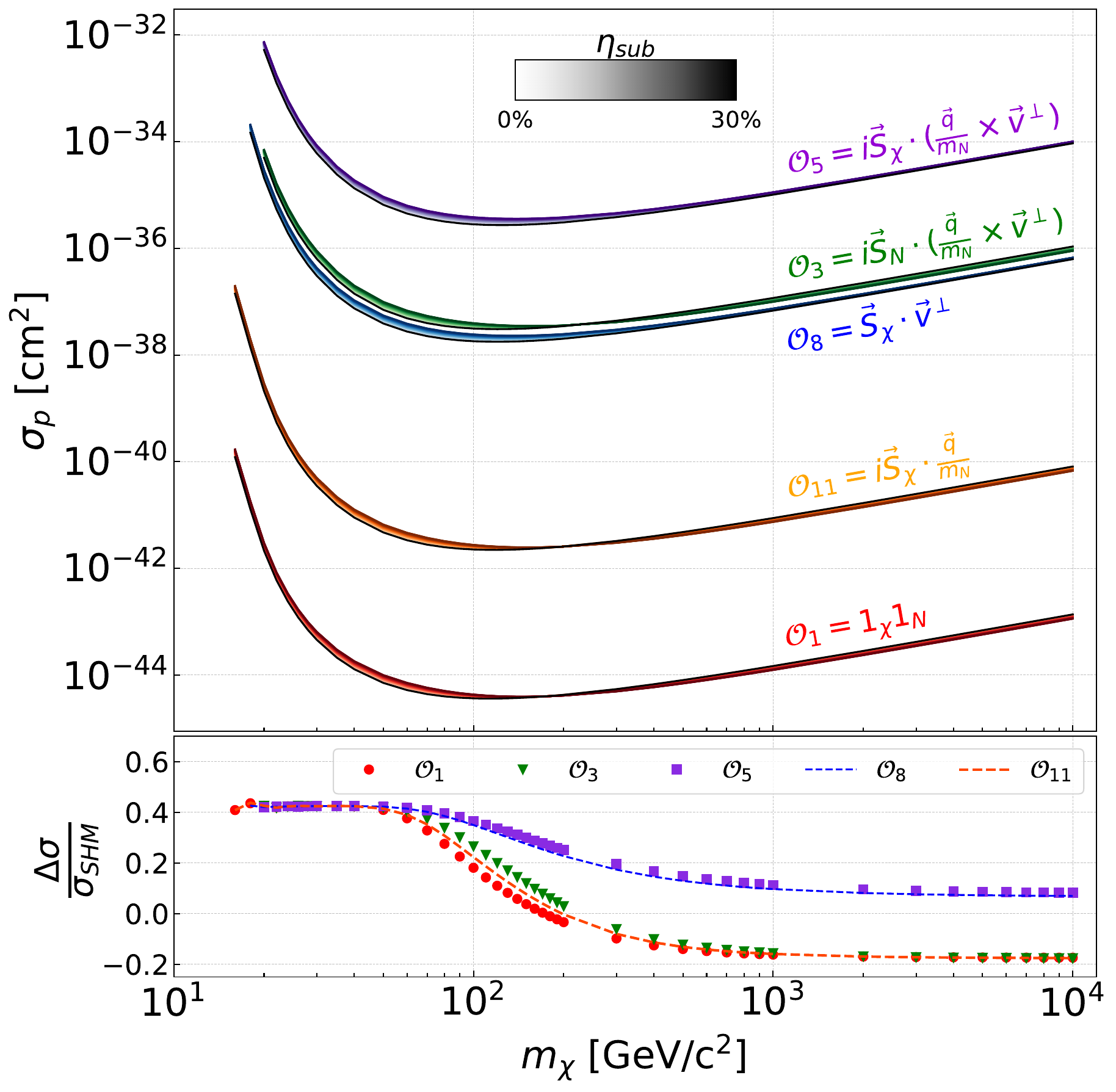}
    \label{fig:limIC1}
  } \\
  \subfigure[G5 streams]{
    \includegraphics[width=0.48\linewidth]{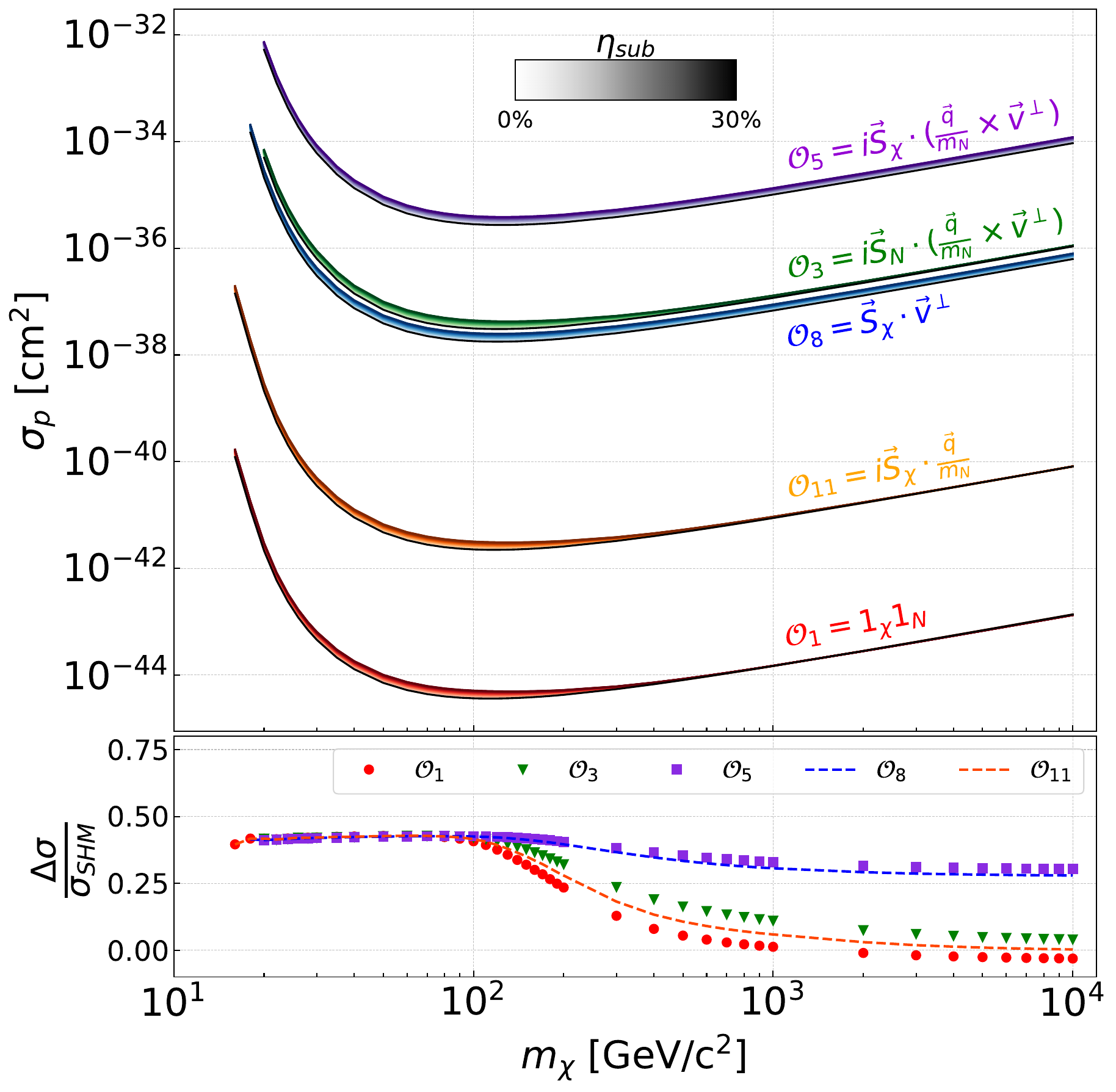}
    \label{fig:limHel}
  }
  \subfigure[G6 streams]{
    \includegraphics[width=0.48\linewidth]{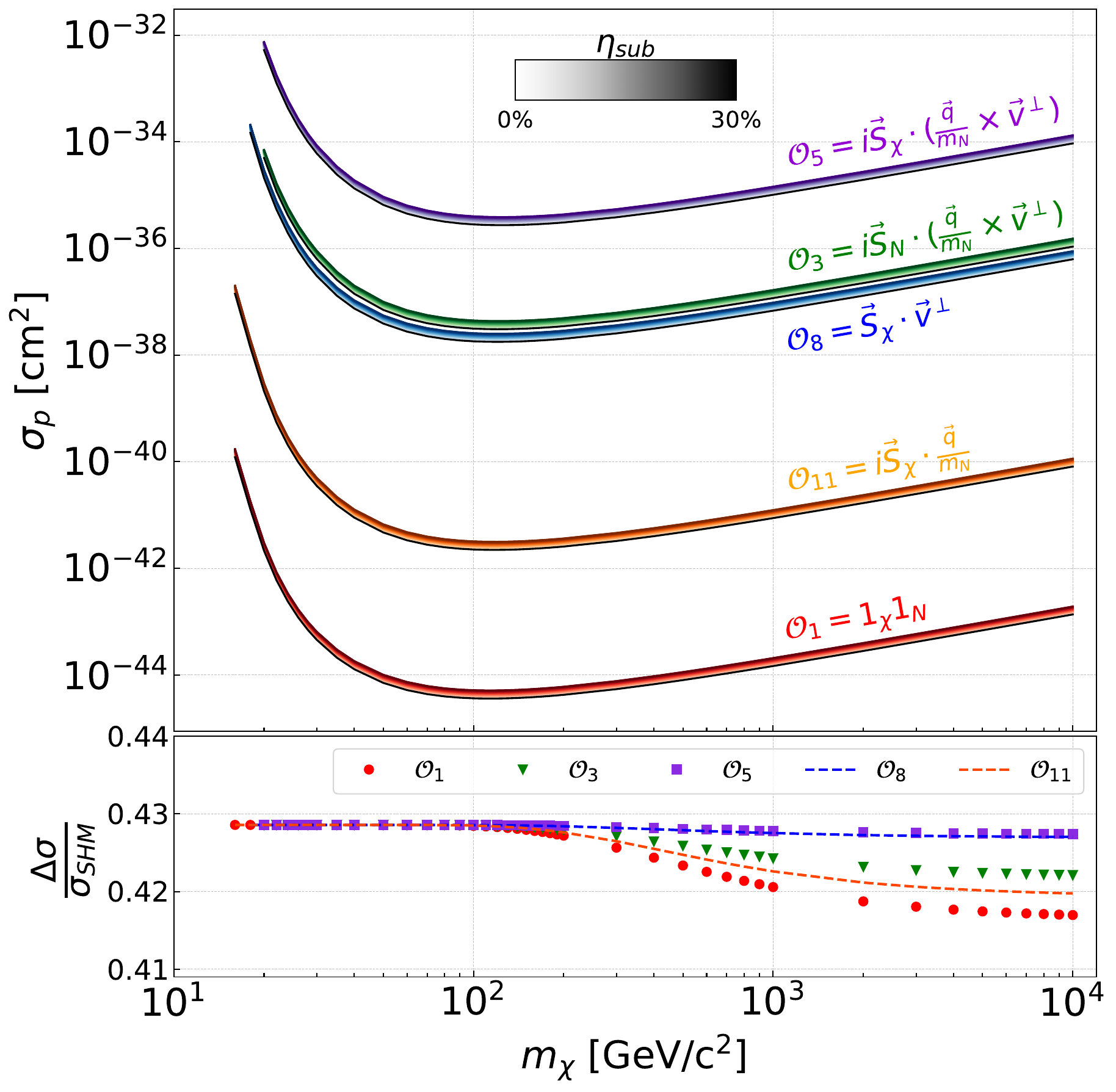}
    \label{fig:limNyx}
  }
  \caption{Upper limits (\NinetyPerCentCL) on DM-nucleon scattering cross sections for the $\CO_1$, $\CO_{11}$, $\CO_3$, $\CO_8$, and $\CO_{5}$ effective operators, in the presence of VDFs corresponding to the G3, G4, G5 and G6 streams, with \SubstructureFraction\ of the DM contained in the specified substructure. Beneath each set of exclusion curves is the relative deviation of each operator with the given substructure at its maximum value compared to the SHM and where $\Delta\sigma$ = $\sigma_\text{sub} - \sigma_\text{SHM}$. } 
  \label{fig:exclusions2}
\end{figure*}

Figs.~\ref{fig:exclusions1} and~\ref{fig:exclusions2} show IS exclusion curves for each NREO under consideration, with each substructure listed in Tab.~\ref{tab:vdfs} varied over the range of \SubstructureFraction.
The relative differences between the exclusion curves drawn with \SubstructureFraction\ at its maximum value and minimum value (corresponding to the SHM) are also shown.

As noted above, DM with  \WIMPMassSymbol$<$\WIMPMassHundredGeV\ exhibit the most sensitivity to substructures, since potential signals in the energy region of interest come from high-velocity tails of the VDFs, where the DM speed can compensate for the lower mass.
For $\CO_1$, $\CO_3$, and $\CO_{11}$, exclusion curves drawn for higher-mass DM become relatively insensitive to most of the substructures considered here, typically deviating from the SHM result by \SI{10}{\percent} or less.

For $\CO_5$ and $\CO_8$, for which $R_k \propto \vperpsquare$, these differences persist at higher \WIMPMassSymbol, as the velocity enhancement of the cross section is independent of $m_\chi$.
As a result, these operators are more sensitive to changes in the VDF than the others are.

Both models of the \GaiaSausage\ result in weaker constraints, due to its relatively low velocity in the laboratory frame.
However, the parametrization by O'Hare \etal~\cite{OHare:2019qxc} (Fig.~\ref{fig:limGSOH}) affects the constraints more strongly at higher \WIMPMassSymbol\ compared to the model by Necib \etal~\cite{necib_inferred_2019} (Fig.~\ref{fig:limGSLN}).
At \WIMPMassThreeTeV, the model by O'Hare \etal\ increases the upper limit by a factor of \SI{2.0}{}, while the model by Necib \etal\ increases it by a factor of \SI{1.7}{}.

However, their relative effects reverse at lower masses.
At \WIMPMassFortyGeV, the model from Necib~\etal\ increases the limit by a factor of \SI{4.7}{}, compared to a factor of \SI{3.1}{}, following O'Hare~\etal.
This behavior is due to the fact that the model described by O'Hare \etal\ is both slower and narrower than the model by Necib~\etal.

The fastest streams, G1 and G2 (Figs.~\ref{fig:limKop} and~\ref{fig:limS1}), strengthen limits the most, with more significant changes for $\CO_5$ and $\CO_8$.
The slowest stream, G6  (Fig.~\ref{fig:limNyx}), decreases sensitivity uniformly across all masses.
DM particles in these substructures have too little kinetic energy across at all considered masses, and so cannot produce a signal in the energy region of interest.
Instead, all candidate signals would come from the residual SHM-like component.

Streams described by G5 (Fig.~\ref{fig:limHel}) consistently yield limits within \SI{40}{\percent} of those obtained from the pure SHM at \WIMPMassFortyGeV, and agree with the SHM prediction to within \SI{3}{\percent} at \WIMPMassThreeTeV.
These streams have a mean close to that of the SHM, and their impact on DM sensitivity mostly derives from the effect of narrowing the VDF.

Limits from streams G3 and G4 are shown in Figs.~\ref{fig:limIC1} and~\ref{fig:limIC2}, respectively.
Both streams decrease sensitivity by up to \SI{40}{\percent} at \WIMPMassFortyGeV, with varying behavior at higher masses.
For $\CO_5$ and $\CO_8$ at higher masses, G4 decreases the sensitivity by up to \SI{9}{\percent}, while it increases the sensitivity by up to \SI{20}{\percent} for the other operators.
At these masses, G3 streams increase the sensitivity for all operators, though limits for $\CO_5$ and $\CO_8$ are strengthened by \SI{24}{\percent}, while the others are improved by up to \SI{20}{\percent}.
These streams increase the sensitivity in some mass ranges and decrease it at other masses due to their narrow VDFs: 
while the VDFs have a slightly higher means than the SHM, their lower spread decreases the population of the high-velocity tail.

\subsection{Simultaneous effects of all model variations}
\label{subsec:allthemodels}    
\begin{figure*}[htb]
\centering
  \subfigure[\WIMPMassSymbol=\WIMPMassFortyGeV]{
    \includegraphics[width=0.48\linewidth]{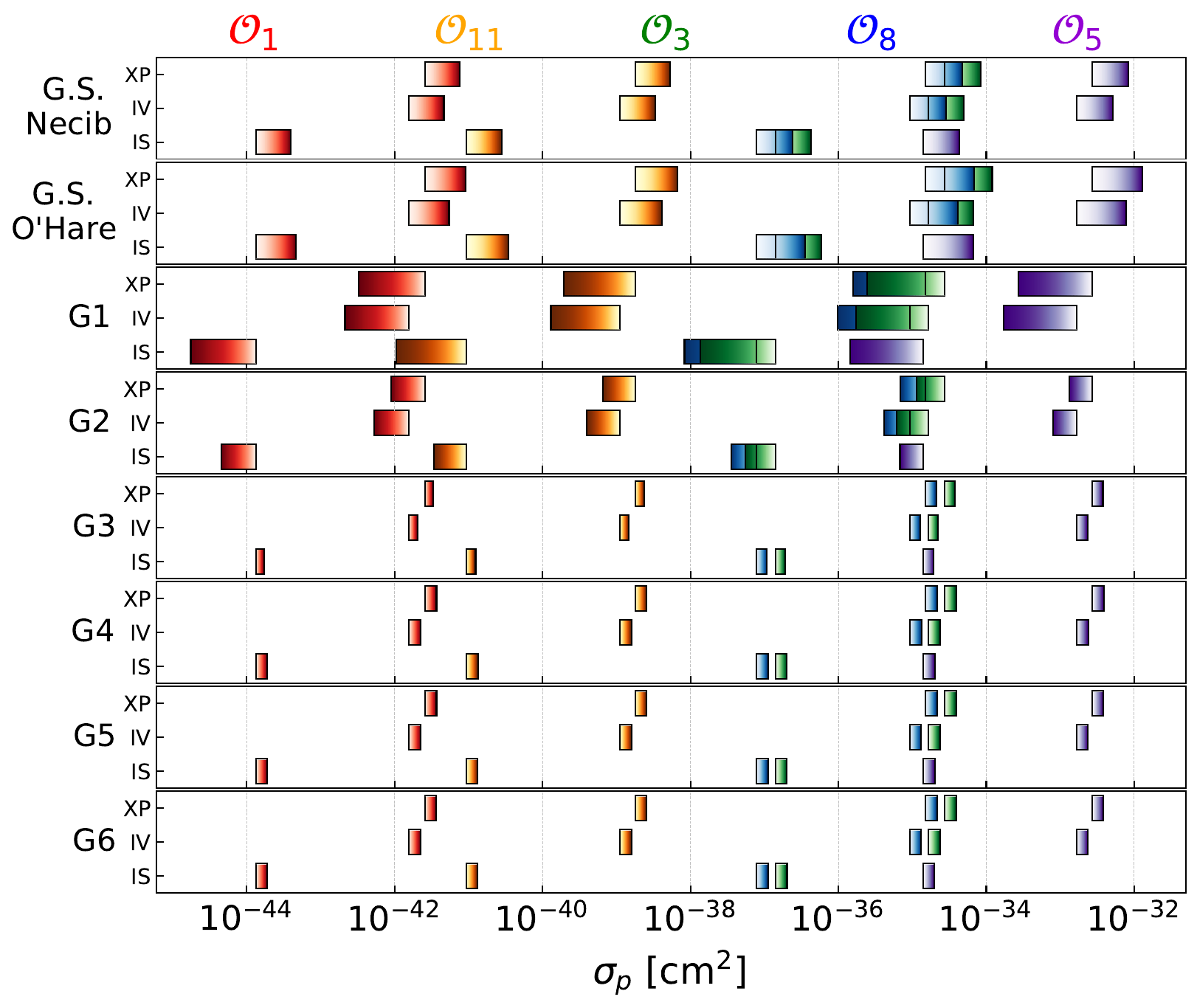}
    \label{fig:limsum40G}
  }
  \subfigure[\WIMPMassSymbol=\WIMPMassHundredGeV]{
    \includegraphics[width=0.48\linewidth]{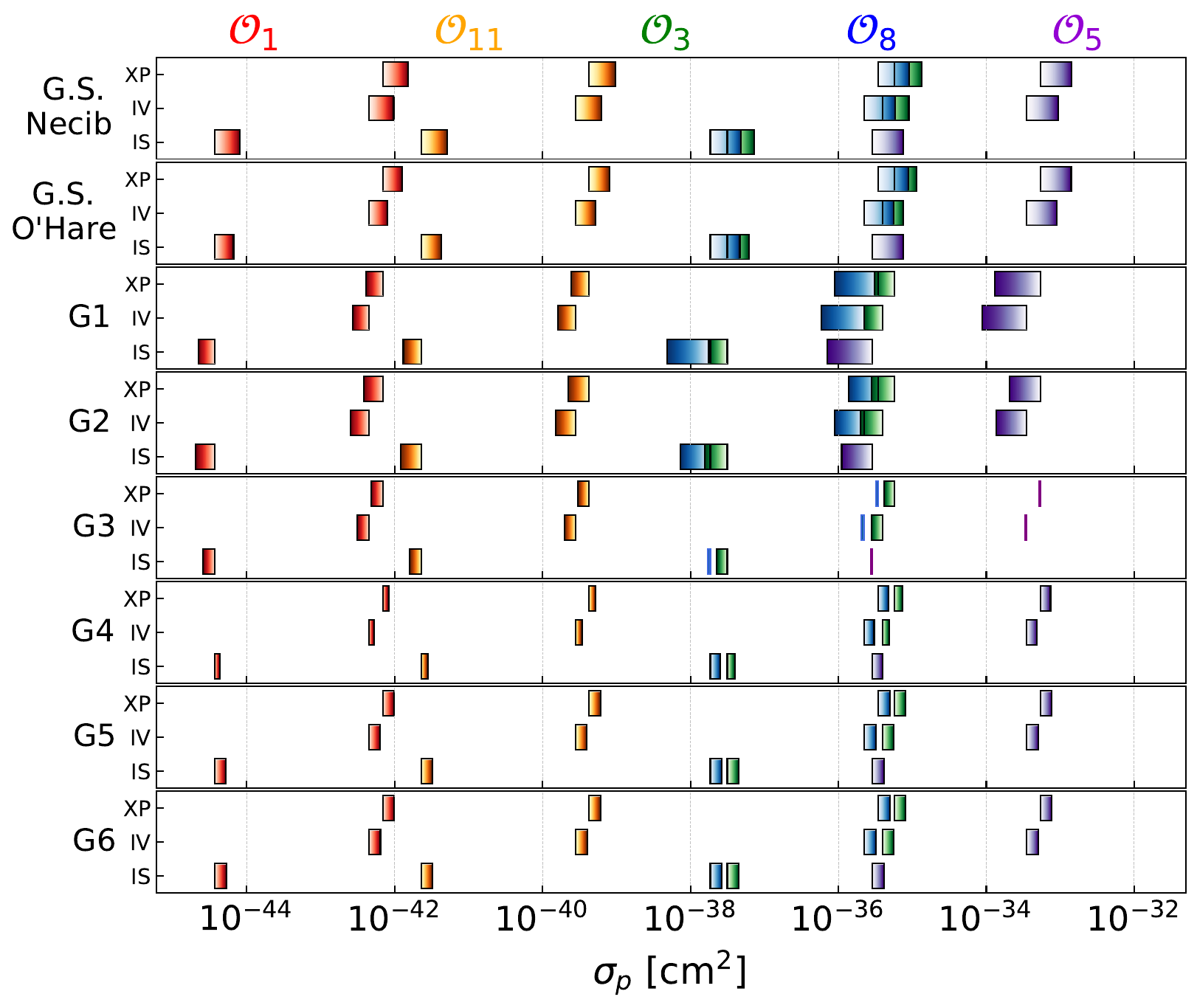}
    \label{fig:limsum100G}
  }
  \\
  \subfigure[\WIMPMassSymbol=\WIMPMassThreeTeV]{
    \includegraphics[width=0.48\linewidth]{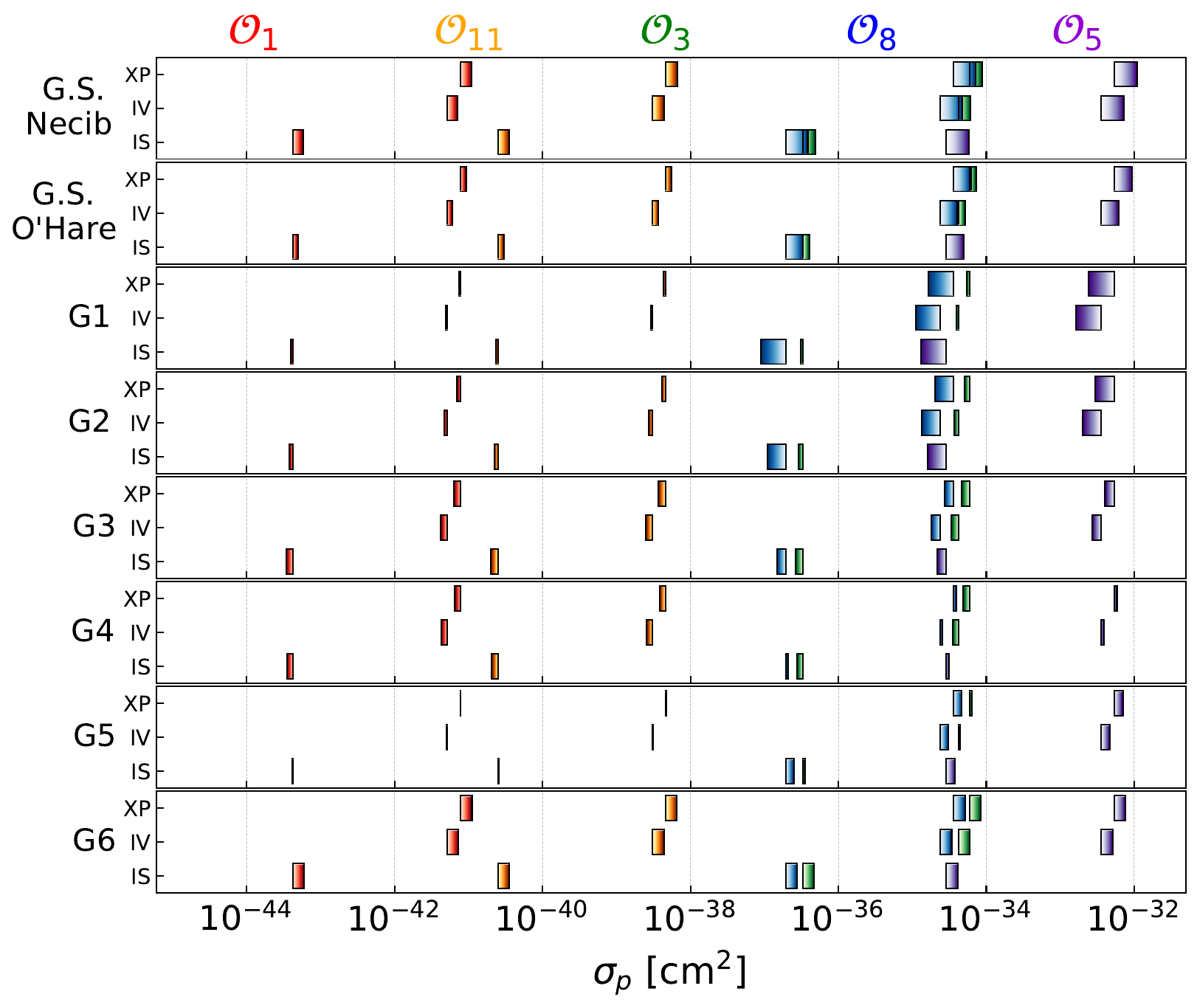}
    \label{fig:limsum3T}
  }

\caption{Summary plots showing upper limits (\NinetyPerCentCL) on the DM-nucleon scattering cross section values at \WIMPMassSymbol=\WIMPMassFortyGeV, \WIMPMassHundredGeV, and \WIMPMassThreeTeV\ for each substructure and isospin scenario. Operators' color (from left to right on the $\sigma_p$-axis) are: $\mathcal{O}_1$ (red), $\mathcal{O}_{11}$ (orange), $\mathcal{O}_3$ (green), $\mathcal{O}_8$ (blue) and $\mathcal{O}_5$(purple). 
Limits labeled ``G.S.'' correspond to the \GaiaSausage\ models by Necib~\etal~\cite{necib_inferred_2019} and by O'Hare~\etal~\cite{OHare:2019qxc}.
The shading in each rectangle indicates the value of \SubstructureFraction, with darker colors denoting higher values. The minimum $\SubstructureFraction = \SI{0}{\percent}$ coincides with the limit constrained with the SHM. }
\label{fig:summaryplots}
\end{figure*}

Limits from all model variations discussed in Sec.~\ref{sec:models} are summarized at three fixed masses, shown in Fig.~\ref{fig:summaryplots}:
Fig.~\ref{fig:limsum40G} shows limits set for \WIMPMassSymbol=\WIMPMassFortyGeV; Fig.~\ref{fig:limsum100G} for \WIMPMassSymbol=\WIMPMassHundredGeV; and Fig.~\ref{fig:limsum3T} for \WIMPMassSymbol=\WIMPMassThreeTeV.

These figures show the \NinetyPerCentCL\ upper limits on the DM-nucleon scattering cross section at each mass, for all operator and isospin scenarios considered.
Upper limits are shown for the VDF groupings presented in Tab.~\ref{tab:vdfs}.
Each rectangle in these figures shows the cross section excluded as the fraction of DM in the substructure \SubstructureFraction\ varies within its specified range, with darker shadings corresponding to higher values of \SubstructureFraction.

The general trends discussed earlier are evident in Fig.~\ref{fig:summaryplots}.
For all operators, constraints on lower-mass DM candidates are most strongly affected by substructures.
Upper limits derived from $\CO_1$, $\CO_3$, and $\CO_{11}$ become relatively insensitive to substructures at higher masses, while $\CO_5$ and $\CO_8$ remain sensitive.

Operators that introduce a factor of $q^2$ or $q^4$ to the DM response function, such as $\CO_3$, $\CO_5$, and $\CO_{11}$ change the shape of the recoil energy spectrum, compared to $\CO_1$.
Similarly, the dependence of $\CO_3$ on $\Phi^{\prime\prime}$ rather than $M$ and variations in the isospin symmetry assumptions change the momentum dependence of the nuclear response function.
Upper limits for interactions with these altered response functions increase in sensitivity by up to \SI{10}{\percent} more when fast substructures are introduced compared to $\CO_1$.

Many of these changes manifest by making the recoil energy spectrum flatter, as discussed in Sec.~\ref{subsec:spectra}. 
Since the current analysis uses the same energy region of interest defined for~\cite{Ajaj:2019imk}, some of these changes occurred at higher energies than were included in this region.
It is therefore likely that extending the analysis region to higher energies will result in limits on these interactions that are more sensitive to substructures.
However, such a study is beyond the scope of the present analysis.

G1 and G2 are the only two substructures that uniformly produce stronger limits for all operators across all masses; the other substructures either always weaken the constraints or have effects that change with mass and operator.
Substructures in this latter category tend to have smaller effects on the constraints compared to others.

The slowest streams, described by G6, uniformly weaken constraints by around \SI{40}{\percent} for all interactions and DM masses.
This constant shift is because most of the DM in these substructures does not have enough kinetic energy to produce a signal in the energy region of interest, and so all potential DM signals must come from the SHM-like component of the VDF, which decreases with \SubstructureFraction.

\subsection{Limits on photon-mediated interactions}
\label{subsec:specificinteractionconstraints}

Limits on photon-mediated interactions are derived using a set of effective operators, as described in Sec.~\ref{subsec:longrange}.

\begin{figure}[!htb]
    \centering
    \includegraphics[width=\linewidth]{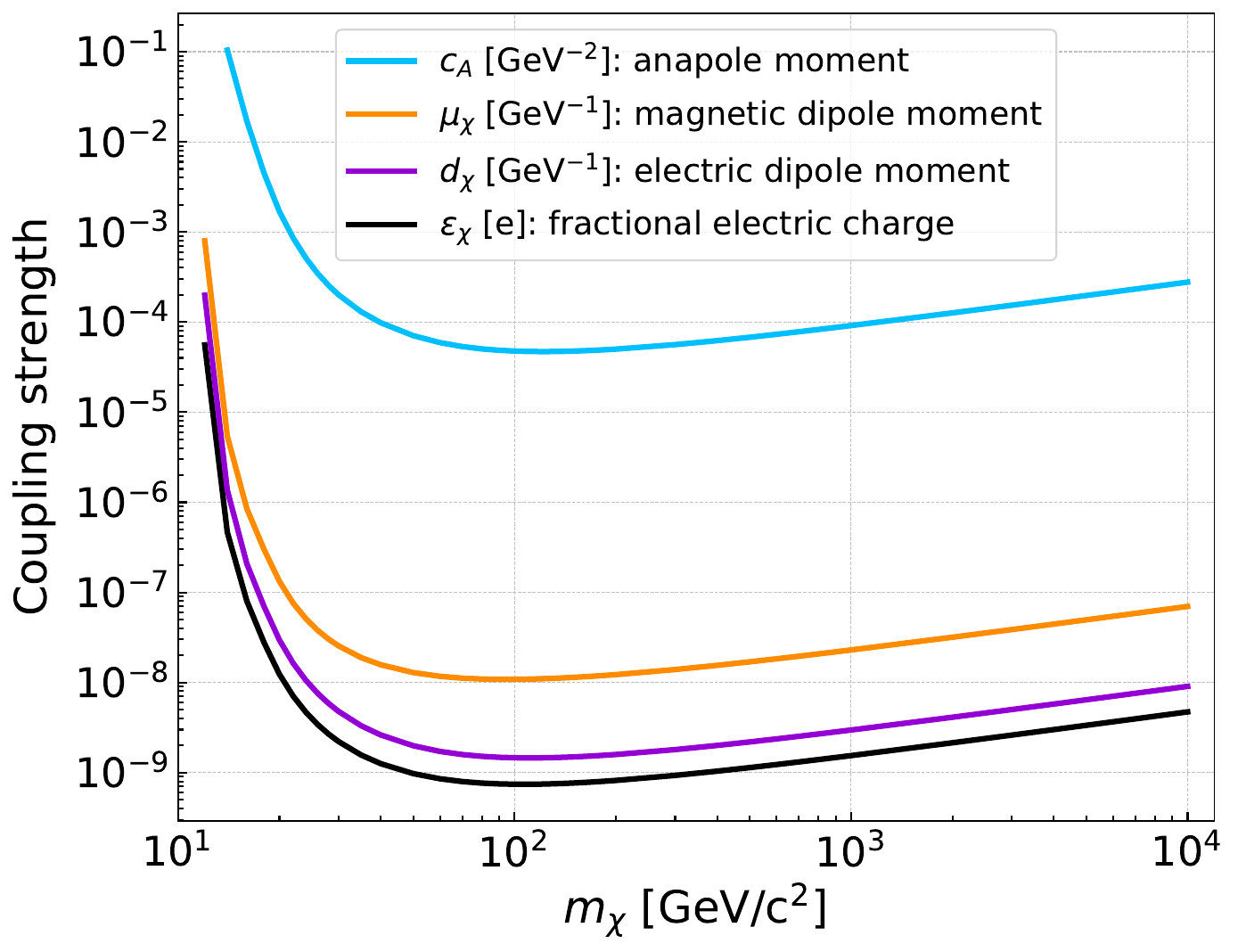} 
    \caption{Exclusion curves on the coupling strength of photon-mediated interactions: anapole, magnetic dipole, electric dipole and millicharged DM using the SHM. }
    \label{fig:speint}
\end{figure}

Upper bounds on the coupling strength of these interactions are shown in Fig.~\ref{fig:speint}, assuming the SHM. 
For \WIMPMassSymbol=\WIMPMassHundredGeV, this analysis excludes an anapole coupling strength $c_A$\SI[per-mode=reciprocal]{>4.8e-5}{\per\square\GeV}, a magnetic dipole $\mu_\chi$\SI[per-mode=reciprocal]{>1.1e-8}{} GeV$^{-1}$, an electric dipole moment $d_\chi$\SI[per-mode=reciprocal]{>1.5e-9}{} GeV$^{-1}$, and an electric charge $\epsilon$\SI{>7.4e-10}{}$e$.

\section{\label{sec:conclusion}Conclusion}

This study provides detailed analyses of \DEAP's constraints on DM-nucleon couplings beyond the standard assumptions of a constant, isoscalar, spin-independent cross section and a Maxwell-Boltzmann DM velocity distribution. 
Using a total exposure of \PaperTwoExpo s, upper limits were placed on the $\CO_1$, $\CO_3$, $\CO_5$, $\CO_8$, and $\CO_{11}$ effective operators in \textit {isoscalar} (IS), \textit {isovector} (IV), and \textit {xenonphobic} (XP) isospin symmetry scenarios, using the NREFT framework described in~\cite{fitzpatrick_effective_2013} and exploring the effects  of various kinematically distinct halo substructures, which are motivated by recent astronomical observations.

Constraints on operators proportional to $v_\perp^n$ are weaker than those proportional to $q$ raised to the same power, which in turn are weaker than constant couplings.
Limits for interactions proportional to the $\Phi^{\prime\prime}$ multipole operator are weaker than those for comparable interactions with $M$, consistent with findings in~\cite{darkside-50_collaboration_effective_2020}.

As shown in~\cite{Yaguna:2019llp}, limits on XP couplings above \WIMPMassHundredGeV\ for \DEAP\ are stronger than those placed by \Xenon\ based on rescaling arguments.
The NREFT framework yields different recoil energy spectra for IS and IV couplings and their cross terms.
In the case of IV interactions, these changes result in slightly weaker limits than are derived using the rescaling method in~\cite{Yaguna:2019llp}, while they comparatively strengthen constraints for XP interactions.

Exclusion curves may substantially change in the presence of kinematically distinct halo substructures.
While these effects are strongest for DM masses at the lower range of \DEAP's sensitivity, they remain particularly strong at higher masses for operators that depend on \vperp.
Constraints on interactions are most significantly strengthened by fast substructures, like the the S1~\cite{OHare:2019qxc} stream and the streams identified by Koppelman \etal~\cite{Koppelman_2018}; constraints are most significantly weakened by slow streams like Nyx~\cite{Necib:2019zbk}, prograde in-falling clumps, and the \GaiaSausage\ debris flow~\cite{necib_inferred_2019,OHare:2019qxc}.

While Nyx is slower than the \GaiaSausage, the potentially high \SubstructureFraction\ for the latter substructure allows it to have a stronger effect.
Both realizations of the \GaiaSausage\ considered here show qualitatively similar effects on upper limits; however, the model described in~\cite{necib_inferred_2019} by Necib~\etal, has stronger effects at lower masses, while the model in~\cite{OHare:2019qxc} by O'Hare~\etal, is more significant at higher masses.
Upper limits set with these models may disagree with each other by around \SI{30}{\percent}.

Limits placed on \vperp-dependent operators were the most sensitive to substructures, while operators proportional to $q$ responded similarly to $\CO_1$. 
However, Figure~\ref{fig:rateVDFs_O1O8O11} shows that the recoil energy spectra for $q$-dependent operators diverge significantly from $\CO_1$ spectra in the presence of substructures at higher energies than were considered for the region of interest.
This observation indicates that greater sensitivity to substructures may arise in searches that extend out to higher energies, up to around \SI{200}{\keV}.

The large variation seen in these limits highlights the importance of the local DM kinetic distribution as a source of uncertainty in the exclusion or discovery of the particle nature of DM.
These effects may be further exacerbated by the presence of multiple substructures.
As demonstrated in~\cite{Buch:2019aiw}, substructures like the \GaiaSausage\ may introduce significant uncertainties in interpreting potential DM detection signals, as well.
However, \cite{Buch:2019aiw} shows that these degeneracies can be resolved by comparing results between experiments, emphasizing the importance of DM searches with different target nuclei.
Halo substructures also have different effects on the recoil energy spectra expected for each operator, potentially allowing spectral information to further resolve these uncertainties.

Kinematic substructures with higher velocities than those discussed here may strengthen these effects. For example, interactions between the Milky Way and the Large Magellanic Cloud (LMC) may result in local substructures with velocities faster than the galactic escape speed if the DM particles originated in the LMC or were accelerated by it. Such substructures are discussed in~\cite{besla_highest-speed_2019}.

Further assessing how various particle and astrophysical models can be resolved is left to future work.
These studies will benefit from the several ongoing efforts to better understand the kinematics of the local DM halo.

Exclusion curves for all operators discussed in the current analysis evaluated for each VDF, including the specific interactions, are available at \eftlink~\cite{deapdataset_v5764847}. Data needed to reproduce the VDFs and recoil energy spectra shown in Figs.~\ref{fig:allvdf}--\ref{fig:rate_vdf_isop} are available there, as well.

\section*{Acknowledgments}

We are grateful to Lina Necib and Bradley J. Kavanagh for valuable input and clarifications, and to Mariangela Lisanti for useful discussions.

We thank the Natural Sciences and Engineering Research Council of Canada,
the Canadian Foundation for Innovation (CFI), 
the Ontario Ministry of Research and Innovation (MRI), 
and Alberta Advanced Education and Technology (ASRIP), Queen's University, 
the University of Alberta, 
Carleton University, 
the Canada First Research Excellence Fund, 
the Arthur B.~McDonald Canadian Astroparticle Phyiscs Research Institute, 
Atomic Energy of Canada Limited's Federal Nuclear Science and Technology Work Plan,
Canadian Nuclear Laboratories,
DGAPA-UNAM (PAPIIT No. IN108020) and Consejo Nacional de Ciencia y Tecnolog\'{\i}a (CONACyT, Mexico, Grant No. A1-S-8960), 
the European Research Council (ERC StG 279980), 
the UK Science and Technology Facilities Council (STFC) (ST/K002570/1 and ST/R002908/1), 
the Russian Science Foundation (Grant No 16-12-10369), 
the Leverhulme Trust (ECF-20130496),
the Spanish Ministry of Science, Innovation and Universities (FPA2017-82647-P grant and MDM-2015-0509),
and the International Research Agenda Programme AstroCeNT (MAB/2018/7)
funded by the Foundation for Polish Science (FNP) from the European Regional Development Fund.
Studentship support by the STFC is acknowledged.
We would like to thank SNOLAB and its staff for support through underground space and logistical and technical services. 
SNOLAB operations are supported by the CFI and the Province of Ontario MRI, 
with underground access provided by Vale at the Creighton mine site.
We thank Vale for their continuing support, including the work of shipping the acrylic vessel underground.
We gratefully acknowledge the support of Compute Canada, 
the Centre for Advanced Computing at Queen's University, 
and the Computational Centre for Particle and Astrophysics (C2PAP) at the Leibniz Supercomputer Centre (LRZ) 
for providing the computing resources required to undertake this work.

\bibliographystyle{deap}
\bibliography{deap}

\end{document}